\documentclass[prd,superscriptaddress,twocolumn,showpacs,nofootinbib]{revtex4}

\usepackage{amsfonts}
\usepackage{amsmath}
\usepackage{amssymb}
\usepackage{bm}
\usepackage{dcolumn}
\usepackage{epsfig}
\usepackage{graphicx}
\usepackage{graphics}
\usepackage[latin1]{inputenc}
\usepackage{latexsym}
\usepackage{rotating}
\usepackage{hyperref}
\usepackage{enumerate}

\newcommand\be{\begin{equation}}
\newcommand\ba{\begin{eqnarray}}
\newcommand\ee{\end{equation}}
\newcommand\ea{\end{eqnarray}}

\newcommand{\msun}{{M}_\ensuremath{\odot}}

\newcommand{\ST}{{\mbox{\tiny ST}}}
\newcommand{\mat}{{\mbox{\tiny mat}}}
\newcommand{\GW}{{\mbox{\tiny GW}}}

\newcommand{\CS}{{\mbox{\tiny CS}}}
\newcommand{\MG}{{\mbox{\tiny MG}}}
\newcommand{\BD}{{\mbox{\tiny BD}}}
\newcommand{\GR}{{\mbox{\tiny GR}}}
\newcommand{\ppe}{{\mbox{\tiny ppE}}}

\newcommand{\tail}{{\mbox{\tiny tail}}}
\newcommand{\PN}{{\mbox{\tiny PN}}}

\newcommand{\pont}{{\,^\ast\!}R\,R}

\newcommand{\IM}{{\mbox{\tiny IM}}}
\newcommand{\MRD}{{\mbox{\tiny MRD}}}
\newcommand{\I}{{\mbox{\tiny I}}}
\newcommand{\M}{{\mbox{\tiny M}}}

\newcommand{\RD}{{\mbox{\tiny RD}}}

\begin{document}
\title {Fundamental Theoretical Bias in Gravitational Wave Astrophysics \\ and the Parameterized Post-Einsteinian Framework}

\author{Nicol\'as Yunes}
\affiliation{Department of Physics, Princeton University, Princeton, NJ 08544, USA.}

\author{Frans Pretorius}
 \affiliation{Department of Physics, Princeton University, Princeton, NJ 08544, USA.}

\date{\today}

\begin{abstract}
  
We consider the concept of {\emph{fundamental bias}} in gravitational wave astrophysics
as the assumption that general relativity is the correct theory of gravity during the entire 
wave-generation and propagation regime. 
Such an assumption is valid in the weak field, as verified 
by precision experiments and observations, but it need not hold in the dynamical strong-field regime where tests are lacking.
Fundamental bias can cause systematic errors in the detection and parameter estimation of signals, 
which can lead to a mischaracterization of the universe through incorrect inferences about source event rates
and populations.
We propose a remedy through the introduction of the {\emph{parameterized post-Einsteinian framework}}, 
which consists of the enhancement of waveform templates via the inclusion of post-Einsteinian parameters.
These parameters would ostensibly be designed to interpolate between templates constructed in general relativity and 
well-motivated alternative theories of gravity, and also include extrapolations that follow sound theoretical principles, such as consistency with conservation laws and symmetries. 
As an example, we construct parameterized post-Einsteinian templates for the binary coalescence of equal-mass, 
non-spinning compact objects in a quasi-circular inspiral. 
The parametrized post-Einsteinian framework should allow matched filtered data to select a specific set of post-Einsteinian parameters
without {\emph{a priori}} assuming the validity of the former, 
thus either verifying general relativity or pointing to possible dynamical strong-field deviations.   

\end{abstract}

\pacs{04.80.Cc,04.80.Nn,04.30.-w,04.50.Kd}
\maketitle

\section{Introduction}
\subsection{The concept of bias and \\ the validity of General Relativity}

The problem of bias has had a negative impact on many scientific endeavors
by causing scientists to draw incorrect inferences from valid 
data. As the next generation of gravitational wave (GW) detectors~\cite{LIGO,VIRGO,GEO} 
promises to open a new observational window onto the universe, one
wonders what bias might be present in this new branch
of astronomy that could lead to incorrect conclusions about the
universe seen in GWs. This is a particularly
pertinent question to address early on, as the first observations
are expected to have quite low signal-to-noise ratio (SNR).

Bias is here to be understood as some set of {\emph{a priori}} assumptions, prejudices or preconceptions 
that affect conclusions derived from properly collected data. We purposely exclude from this definition 
experimentally-induced systematic errors, as we wish to concentrate on errors introduced 
during post-processing and analysis. In GW astrophysics, a principal 
source of bias is the {\emph{a priori}} assumption, often unstated, that general relativity (GR) 
is the correct theory that describes all gravitational phenomena at the scales 
of relevance to GW generation and propagation. This bias is ingrained in many of the
detection and parameter estimation tools developed to mine GW data, and extract
astrophysical information from them (see for example~\cite{Fairhurst:2009qm}
and references therein).

The study of theoretical bias in GW astrophysics is not completely new. Recently, 
Cutler and Vallisneri~\cite{Cutler:2007mi} analyzed the issue of systematic error generated by the use
of inaccurate template families. This issue can be broadly thought of as a {\emph{modeling bias}},
where the preconception relates to physical assumptions to simplify the solutions considered (e.g. that all
binaries have circularized prior to merger), or unverified assumptions about the accuracy
of the solution used to model the given event.
The study in ~\cite{Cutler:2007mi} considered inaccuracies arising from errors in solutions 
to the Einstein equations due to the use of the post-Newtonian (PN) approximation scheme,
and they found these inaccuracies could dominate the error budget. 
This paper differs from that analysis in that we are concerned with the validity of the Einstein 
equations themselves, an assumption that can be considered a {\emph{fundamental bias}}. 

Systematic errors created by fundamental bias may be as large, if not larger, than those induced by 
modeling bias, as waveforms could deviate from the GR prediction dramatically in the dynamical strong-field,
if GR does not adequately describe the system in that region. This is particularly worrisome for template-based
searches, as the event that will be ascribed to a detection will be the member of the template
bank with the largest SNR. Given that
GR is quite well tested in certain regimes, many sources cannot have deviations so far from GR 
as to prevent {\em detection} with GR templates (albeit with lower SNR). Thus, if templates are used based
solely on GR models, although the corresponding events may be ``heard'', any unexpected information the signals may
contain about the nature of gravity will be filtered out.

But is it sensible to expect GR deviations at all? GR has been tested to high accuracy in the Solar System
and with binary pulsars~\cite{lrr-2006-3,Will:1993ns}. However, in the Solar System gravitational fields are weak and particle 
velocities are small relative to the speed of light, and all such tests only probe linear perturbations in 
the metric beyond a flat Minkowski background. Binary pulsar tests begin to probe the strong-field regime in terms
of the {\em compactness} of the source ({\emph{i.e.~}}the ratio of the source mass to its 
radius), though known systems are still quite weak with regard to the strength of the dynamical gravitational 
fields (characterized by the ratio of the total mass of the system to the binary separation).
In this sense then, the {\em dynamical, strong-field} region of GR has so far eluded direct
observational tests~\cite{Psaltis:2008bb}.

The lack of experimental verification of GR in the dynamical strong-field could be remedied through GW observations. 
For example, a compact object such as a black hole (BH) or neutron star (NS) spiraling into
a supermassive BH will emit waves that carry a map of the gravitational field of the
central BH, a program of relevance to the planned {\emph{Laser Interferometer Space Antenna}} mission 
(LISA)~\cite{lisa}, and sometimes referred to as
{\emph{bothrodesy}}~\cite{Ryan:1997hg,Collins:2004ex,Hughes:2004vw,Schutz:2009tz}.
Such maps would then allow for detailed tests of alternative theories of gravity. Additional
work on constraining alternative theories have
concentrated on using either the binary inspiral or the post-merger ringdown phase.
With the former, studies have focused on the search for
the graviton Compton wavelength~\cite{Will:1997bb,Finn:2001qi,sutton:2002:bgm,Will:2004xi,Berti:2005qd,Berti:2004bd,Arun:2009pq,Stavridis:2009mb}, 
the existence of a scalar component to the gravitational interaction \cite{Damour:1998jk,Scharre:2001hn,Will:2004xi,Berti:2005qd,Berti:2004bd,Yagi:2009zm,2009arXiv0907.2186C,Stavridis:2009mb,Arun:2009pq} and
the existence of gravitational parity violation~\cite{Alexander:2007:gwp,Yunes:2008bu}.
Studies of the ringdown phase have concentrated on violations of the GR no-hair theorem (that the mass and spin
completely determine the gravitational field of a rotating BH) by proposing
to verify certain consistency relations 
between quasinormal ringdown (QNR) modes~\cite{Thorne:1997cw,Dreyer:2003bv,Berti:2005ys,Berti:2006qt}.  
All of the above tests, however, have focused on specific alternative theories and have not investigated the issue of fundamental
bias that we wish to address here. 

\subsection{Toward a framework for detecting fundamental bias in GW observations of compact object mergers}

In this paper, we are not so much interested in {\em specific} tests of {\em particular}
alternative theories, but rather we want to introduce a framework that may allow one to quantifiably investigate
the consequences of fundamental bias in GW astronomy. We propose to do so via the introduction
of the {\emph{parameterized post-Einsteinian}} (ppE) framework, analogous to the  parameterized post-Newtonian
(ppN)~\cite{Nordtvedt:1968qs,1972ApJ...177..757W,1971ApJ...163..611W,1972ApJ...177..775N,1973ApJ...185...31W,lrr-2006-3}
or the parameterized post-Keplerian (ppK) ones~\cite{1991STIN...9219818D,1992PhRvD..45.1840D,lrr-2006-3},
where model waveforms are enhanced in a systematic and well-motivated manner by parameters that can measure
deviations from GR. The construction of such a ppE framework is a quixotic task in general, 
and so to make it more manageable we will concentrate on a subset of possible GW sources:
the inspiral, merger and ringdown of binary BH-{\em like} compact objects.

Given that we want to liberate GW astronomy from the assumption that GR is correct in all regimes, what sense then
does it make to begin with a binary BH merger, an event  wholly within the realm of GR?
First, there is strong observational evidence that highly compact objects exist,
and in many cases the observations are consistent with the supposition that
they are BHs~\cite{Narayan:2005ie}: the high luminosity of quasars
and other active galactic nuclei (AGN) can be explained by gravitational binding energy released through gas accretion onto
supermassive ($10^6-10^9 \msun$) BHs at the centers of the galaxies~\cite{Rees:1984si,Ferrarese:2004qr}; several dozen X-ray binary
systems discovered to date have compact members too massive to be NSs, and exhibit phenomena consistent with
matter interactions originating in the strong-field regime of an inner accretion disk~\cite{McClintock:2003gx};
the dynamical motions of stars and gas about the centers of nearby galaxies
and our Milky Way Galaxy suggest the presence of very massive, compact
objects, the most plausible explanation being supermassive
BHs~\cite{Gebhardt:2000fk,Schodel:2002py,Ghez:2003qj}. Therefore, it is not too bold
to assume that even if GR does not accurately describe compact objects,
whichever theory does must nevertheless still permit BH-{\emph{like}} solutions.
Second, from studies of the orbital decay of the Hulse-Taylor~\cite{Hulse:1974eb} and more recently
discovered binary pulsars~\cite{Burgay:2003jj,Kramer:2006nb}, it is again
a rather conservative conclusion that the early evolution of binary compact
objects in the universe is adequately governed by GR---namely, that binary systems
are unstable to the emission of quadrupole gravitational radiation.

Based on the above considerations, starting with GR binary BH merger waveforms seems sound.
We are then faced with the question of how to modify these waveform in a sensible manner.
In theory, there are uncountably many {\em conceivable} modifications
to GR that only manifest in the late stages of the merger. To make this question
manageable, we shall guide our search for ppE expansions by looking to 
alternative theories that satisfy as many of the following criteria as possible:
\begin{enumerate}[(i)]
\item {\emph{Metric theories of gravity}}: theories where gravity is a manifestation
of curved spacetime, described via a metric tensor, and which 
satisfies the weak-equivalence principle~\cite{Misner:1973cw}. 
\item {\emph{Weak-field consistency}}: theories that reduce to GR sufficiently when gravitational fields
are weak and velocities are small, {\emph{i.e.}}~to pass all precision, experimental 
and observational tests.   
\item {\emph{Strong-field inconsistency}}: theories that modify GR in the dynamical strong-field 
by a sufficient amount to observably
affect binary merger waveforms.
\end{enumerate}
Notice that the weak-field consistency criterion also requires the existence and stability of physical solutions, such as
the Newtonian limit of the Schwarzschild metric to describe physics in the Solar System.
One might also wish that other criteria be satisfied, such as well-posedness of the initial-value problem, the existence
of a well-defined, relativistic action, and that the theory be well-motivated from fundamental physics (eg.~string theory,
loop quantum gravity, the Standard Model of elementary interactions, etc), but we shall not impose such additional 
requirements here.

Instead of concentrating on a particular theory that satisfies the above criteria, 
here we are more concerned with being able to measure {\emph{generic}} deviations
from GR predictions. In spite of the possibly uncountable number of modifications one can 
introduce to the GR action, parameterizations of generic deviations is easier if one concentrates
on the waveform observable, the {\emph{response function}}, which in turn can be described by a complex 
function in the frequency domain. Deviations from GR then translate into deviations in the waveform
amplitude and phase. Such deviations could be induced, for example, by modifications 
to the GW emission formulae (e.g. the quadrupole formula), new polarization modes, 
new propagating degrees of freedom and new effective forces.
In order to ease the flow of the paper, we will not further discuss details
of how certain alternative theories modify the gravitational waveform here,
referring the interested reader to Appendix A. 

\subsection{Two key questions related to fundamental bias}\label{sec_two_qs}

In view of the above considerations, we shall attempt to lay the foundations to answer the following 
two questions related to fundamental bias in GW astronomy, here focusing on binary compact object mergers:
\begin{enumerate}[(i)]
\item  Suppose gravity is described by a theory differing from GR in the
dynamical, strong-field regime, but one observes a population of merger events 
filtered through a GR template bank. What kinds of systematic errors and incorrect 
conclusions might be drawn about the nature of the compact object population due to this fundamental bias?  
\item Given a set of observations of merger events obtained with a 
GR template bank, can one quantify or constrain the level of consistency of these observations with
GR as the underlying theory describing these events?
\end{enumerate}

As we will discuss in the next subsection, the ppE framework will be able to answer 
question $(2)$, at least for the class of deviations 
from GR considered here.
As for question $(1)$, we shall not specifically address it in this
paper, though will here give a couple of brief examples to clarify the question.
As a first example, suppose the ``true'' theory of gravity differs from GR in that scalar
radiation is produced in the very late stages of a merger, and during the ringdown.
This will cause the inspiral to happen more quickly compared to GR, resulting in
a late-time dephasing of the waveform relative to a GR template. Also, less power may be
radiated in GWs during the ringdown phase. In all then, these events may be detected
by GR templates, though with systematically lower SNR, {\emph{i.e.}}~be seemingly more distant. 
Here then, the fundamental bias could make one incorrectly infer that merger events were 
more frequent in the past. 

For a second hypothetical example, consider an extreme mass ratio merger, where a small
compact object spirals into a supermassive BH. Suppose that 
a Chern-Simons (CS)-like correction is present, altering the near-horizon
geometry of the BH as described in~\cite{Yunes:2009hc,Konno:2009kg}.
To leading order, the CS correction reduces the effective gravitomagnetic
force exerted by the BH on the compact object; in other words, the GW
emission would be similar to a compact object spiraling into a GR Kerr
BH, but with smaller spin parameter $a$.
Suppose further that near-extremal
($a\approx 1$) BHs are common (how rapidly astrophysical
BHs can spin is an interesting and open question).
Observation of a population
of CS-modified Kerr BHs using GR templates would
systematically underestimate the BH spin, leading to the
erroneous conclusion that near-extremal BHs are uncommon,
which could further lead to 
incorrect inferences about astrophysical BH formation and
growth mechanisms.

\subsection{Toward a ppE Construction}

The concept of a ppE framework is in close analogy to the ppN 
one, proposed by Nordtvedt and Will~\cite{Nordtvedt:1968qs,1972ApJ...177..757W,1971ApJ...163..611W,1972ApJ...177..775N,1973ApJ...185...31W,lrr-2006-3}
to deal with an outbreak of alternative theories of gravity in the 1970's. Such a framework 
allows for model-independent tests of GR in the Solar System, through the 
introduction of ppN parameters in the weak-field expansion of the metric tensor.
When these parameters take on a specific set of values, the metric 
tensor becomes identical to that predicted by GR, while when it
takes on different values, the gravitational field becomes that predicted by certain alternative theories. 
In the same way, we introduce ppE parameters to interpolate between different theories, but 
we parameterize the GW response function instead of the metric tensor, as the former is the observable in GW astrophysics.

We shall follow the ppN route in the 
construction of a ppE framework: we will explore a set of alternative theories and their effects 
on GWs, and from these, we will phenomenologically infer and engineer a ppE template family that not only captures 
the waveforms from known theories, but also other phenomenological corrections. 
As we shall see, our approach reproduces 
corrections to the GW response functions predicted by a large class of alternative 
theories of gravity proposed to date, including Brans-Dicke
theory, massive graviton theories and non-dynamical CS modified gravity.

As a first step, we shall {\emph{not}} consider here the effect of scalar, vectorial or all six tensorial polarizations 
on the response function, nor will we attempt to classify alternative theories based on the different perturbative 
modes that can be excited. 
In fact, such a classification already exists, the $E(2)$ scheme~\cite{Eardley:1973br,Will:1993ns}, which organizes different alternative
theories according to whether they excite certain contractions of the Weyl tensor as characterized by the Newman-Penrose (NP) scalars.
Here we will concentrate on the two ``plus'' and ``cross'' tensorial degrees of freedom,
and how these change in alternative theories.
Additional excited modes will only be considered insofar as they affect the source dynamics, and
thus indirectly the structure of the plus and cross modes. 

In the future it would be useful to extend the ppE templates to include additional GW polarizations and modes.
However that would be a non-trivial effort, the main reason being that currently there is a 
lack of alternative theories that simultaneously satisfy the criteria discussed above and 
allow for the excitation of additional modes. For example, 
Brans-Dicke theory does possess a breathing mode, though it is strongly suppressed by constraints on its coupling constant. 
Similarly, Einstein-Aether theory generically has 5 propagating degrees of freedom, but these decouple and radiation can be
shown to remain quadrupolar~\cite{Jacobson:2008aj}. 
Thus, unlike with the other ppE extensions considered here, there is little theoretical guidance available 
on how GR waveforms would be modified if additional GW polarizations are excited. Without such guidance then,
it will be almost impossible to argue that any proposed extensions are well-motivated.

A separate issue is whether the effect of additional polarization states can be observed with
contemporary or planned detectors.
For direct measurement one requires $2 N$ arms to be sensitive to $N$ polarization states ~\cite{Neil-priv}, for example 
using multiple ground-based detectors, Doppler tracking of multiple spacecraft, or pulsar timing arrays~\cite{Hellings:1978vp}. 
Alternatively, a space-borne detector such as LISA can be sensitive to multiple polarization states for events
that last a sizable fraction of the spacecraft's orbit, as the motion of detector then effectively samples the waveform
with multiple arm orientations.
For ground-based detectors, binary merger events happen too quickly for a similar strategy to be effective.
However, other sources, for example the waves that could be emitted by a ``mountain'' on a pulsar, will produce 
coherent GWs over many rotation periods of the Earth, and thus could also contain information on multiple
polarization states. We will not consider these interesting issues here.
 
\subsection{A ppE template family for the inspiral, merger and ringdown
of black hole-like compact objects}
 
For this initial study, we construct ppE templates describing only the {\emph{quasi-circular}} coalescence of {\emph{non-spinning}} and
{\emph{equal-mass}} compact objects. Even within this restricted class of events, the ppE construction is non-unique,
and certainly more refined versions could be developed.
We introduce several ppE template families that vary in the number of ppE parameters, and thus, in the amount of 
fundamental bias each family assumes in its construction. The frequency-domain (overhead tilde) ppE templates 
with the least number of ppE parameters that we derive is the following (more general representations are 
provided in Sec.~\ref{com-waveform}):
\ba
\tilde{h}(f) =
 \begin{cases}
\tilde{h}^{(\GR)}_{\I}(f)\cdot
\left(1 + \alpha u^{a} \right) e^{i \beta u^{b}}  & \text{$f < f_{\IM}$}, \\
  \gamma u^{c} e^{i (\delta + \epsilon u)}& \text{$f_{\IM} < f < f_{\MRD}$}, \\
   \zeta \frac{\tau}{1 + 4 \pi^{2} \tau^{2} \kappa \left(f - f_{\RD}\right)^{d}} 
& \text{$f > f_{\MRD}$},
\end{cases}
\label{main-eq}
\ea
where the subscript $IM$ and $MRD$ stand for inspiral-merger and merger-ringdown. 
In the inspiral phase ($f < f_{\IM}$), the GW is described by a ``chirping'' complex exponential, 
consisting of the GR component $\tilde{h}^{(\GR)}_{\I}(f)$ corrected by ppE amplitude and phase
functions with parameters $(\alpha, a, \beta, b)$.
Here $u = \pi {\cal{M}} f$ is the inspiral reduced frequency, ${\cal{M}}=M \eta^{3/5}$ is the chirp mass
with symmetric mass ratio $\eta = m_{1} m_{2}/M^{2}$ and total mass $M = m_{1} + m_{2}$
(though again, we will only focus on the equal mass case).
The merger phase ($f_{\IM} < f < f_{\MRD}$) is treated as an interpolating region between
inspiral and ringdown, where the merger parameters $(\gamma,\delta)$ are set by continuity, and the 
merger ppE parameters are $(c, \epsilon)$. In the ringdown phase ($f > f_{\MRD}$), the GW is described
by a single-mode generalized Lorentzian, with real and imaginary dominant frequencies $f_{\RD}$ and $\tau$, ringdown 
parameter $\zeta$ also set by continuity, and ppE ringdown parameters $(\kappa,d)$. The transition
frequencies $(f_{\IM},f_{\MRD})$ can either be treated as ppE parameters, or (for example) set to the GR light-ring frequency 
and the fundamental ringdown frequency, respectively. In a later section, we shall present more general ppE waveforms
with inherently less fundamental bias, some of which are reminiscent to the work of~\cite{Arun:2006yw,Arun:2006hn,2009arXiv0907.2186C}, 
except that here we are interested in tests of GR through the determination of ppE parameters, instead of the measurement of 
PN ones.

The ppE template presented above is {\emph{non-unique}};
in fact, the ppE framework as a whole is inherently non-unique, as are the ppN or ppK ones, 
because a {\emph{finite}} parameterization cannot represent an infinite space of alternative theory templates. 
However, this ppE family is {\em minimal} within
the class of templates considered here, as it employs the smallest number of ppE parameters necessary
to reproduce corrections to the GW response function
from well-known alternative theories of gravity in the inspiral phase: 
\begin{enumerate}[(i)]
\item GR is reproduced with $(\alpha,a,\beta,b) = (0,a,0,b)$, $(c, \epsilon) = (-2/3, 1)$ and 
$(\kappa,d) = (1,2)$;
\item Jordan-Brans-Dicke-Fierz, or simply Brans-Dicke theory (BD), 
with $(\alpha,a,\beta,b) =(0,a,\beta_{\BD},-7/3)$, and $\beta_{\BD}$ 
related to the Brans-Dicke coupling parameter (see Eq.~\eqref{betaBD});
\item Massive graviton (MG) theories with $(\alpha,a,\beta,b) = (0,a,\beta_{\MG},-1)$, and $\beta_{\MG}$ related
to the graviton Compton wavelength (see Eq.~\eqref{betaMG}); 
\item CS modified gravity with $(\alpha,a,\beta,b) = (\alpha_{\CS},1,0,b)$, and $\alpha_{\CS}$
related to the CS coupling parameter (see eg.~Eq.~\eqref{resp-CS}).
\end{enumerate}

The allowed range of the ppE parameters is not completely free, as Solar System and binary pulsar experiments
have already constrained some of them. The subset of inspiral ppE templates that reproduce well-known theories
after fixing $a$ and $b$ is then constrained to some range in $(\alpha,\beta)$. More precisely, since the Brans-Dicke
coupling parameter has been constrained to $\omega_{\BD} >  4 \times 10^{4}$ by the Cassini spacecraft~\cite{Bertotti:2003rm}, 
this automatically forces $\beta_{\BD} < 3.5 \times 10^{-8} (s_{1} - s_{2})^{2} \eta^{2/5}$ when $b=-7/3$, where $s_{1,2}$ are the
sensitivities of the binary components. For example, for a $1.4 M_{\odot}$ binary NS system, $\beta_{\BD} \lesssim 10^{-9}$ when $b=-7/3$. 
Similarly, since the Compton wavelength of a massive graviton has already been constrained to 
$\lambda_{g} > 3.4 \times 10^{15} \; {\rm{km}}$ from pulsar timing observations~\cite{Baskaran:2008za}, this implies 
$\beta_{\MG} < 8.5 \times 10^{-31} D {\cal{M}} (1 + z)^{-1} \; {\rm{km}}^{-2}$ when $b = -1$, where $D$ is a distance measure to the source.
For example, for a $1.4 M_{\odot}$ binary NS system at a redshift of $z=0.1$, one finds the prior
$\beta_{\MG} \lesssim 1.6 \times 10^{-8}$ when $b=-1$. Finally, the non-dynamical CS coupling parameter has been constrained by binary pulsar observations to be such 
that for an aligned binary (zero inclination angle), $\alpha_{\CS} \lesssim 70 \times 10^{-33} D/{\cal{M}}$ when $a = 1$, which for a binary NS 
at redshift $z=0.1$ translates to $\alpha_{\CS} \lesssim 4 \times 10^{-10}$.

One can also consider placing priors on other ppE parameters introduced in Eq.~\eqref{main-eq} without fixing
the waveform to represent a particular well-known theory.
However, since the merger and the ringdown have not been constrained at all by observation, one cannot
really restrict $(c,\epsilon,\kappa,d)$. On physical grounds, one can only require that $d>0$ and real 
and $(\epsilon,\kappa)$ also real, such that the waveform is well behaved at large frequencies. 
Since the quadrupole formula has been verified to leading order with binary pulsar observations, one also
expects that for low-frequencies $\alpha u^{a} < 1$, which for example implies that if $0< a \ll 1$ then $|\alpha| < 1$, 
while if $a \gg 1$ then $\alpha$ is essentially unconstrained. If one does not expect gravitational radiation 
to be sourced below dipolar order one must have $b > -7/3$; similarly, if one expects strong-field,
source generation deviations to only arise beyond quadrupolar order then $b >  -5/3$. 
Of course, the stronger the priors imposed, the stronger the bias the ppE model inherits.

\subsection{The Possible Role of the ppE Framework in Data Analysis}
\label{sec_alg}

We do not propose here to employ the ppE templates for direct detection, but  rather for {\emph{post-detection analysis}}. 
Following the detection of a GW by a pure GR template, that segment of data could then be reanalyzed
with the ppE templates. In principle, one should search over all system and ppE parameters, but in practice, since the ppE waveforms are
deformations of GR templates, one can restrict attention to a neighborhood of the system parameter space centered around the best-fit ones obtained
through filtering with a GR waveform. 

The templates presented above {\em do} contain some modeling error, {\emph{i.e.}}, 
the subset of the ppE template family with all parameters set to the GR 
values---{\em the ppE-GR templates}---are not the ``exact'' solution to the
event in GR (for which there is no known closed form solution). 
What is more important though is that the ppE-GR templates have a high overlap or fitting factor with 
the correct waveform, which is why the ppE templates have been built as generalizations of GR approximations
with that particular property~\cite{Ajith:2007kx}. The modeling error in the ppE-GR templates could be determined by computing
fitting factors between them and the corresponding GR templates, folding in the detector noise curve, or following the formalism 
laid out in~\cite{Cutler:2007mi}. This would then provide a measure of the efficacy of the ppE templates in detecting deviations
from GR, or constraining alternative models, with the given event.

Although the practical implementation of an efficient post-detection analysis pipeline will be relegated to future work, one should 
not necessarily restrict such analysis to a frequentist approach. An alternative would be to perform a Bayesian model selection 
study~\cite{Umstaetter:2007su,cornish:083006} to determine whether the evidence points toward GR or a GR deviation 
(through the non-vanishing of ppE parameters). The idea is that more parameters in a template bank will generically improve the fit to 
the data, even if there is no signal present. Thus, any search should be guided by careful consideration of priors (such as those discussed earlier)
or through the computation of {\emph{evidence}} associated with a given model, for example as in~\cite{Feroz:2009de}. Such considerations
are particularly important for low SNR events, such as those expected for ground-based detector sources.

A clear path then presents itself for possible strategies that one might pursue in the study of fundamental bias:
\begin{enumerate}[(i)]
\item Given a GR signal and a ppE template, how well can the latter extract the former? how much modeling
error is intrinsic in the ppE template family? 
\item Given a non-GR signal and a GR template, how much fundamental bias-induced systematic error is
generated in the estimation of parameters? Can the signal even be extracted?
\item Given a non-GR signal and a ppE template, how well can the latter extract the former? How 
well can intrinsic and ppE parameters be estimated? 
\end{enumerate}
These questions constitute a starting point for future data analysis studies to investigate the issue
of fundamental bias in GW astronomy and refinements of the ppE framework.  

\subsection{Organization of this paper}

An outline of the remainder of this paper is as follows:
Sec.~\ref{anatomy} discusses the anatomy of binary BH coalescence in GR;
Sec.~\ref{Inspiral-Sec} derives some model-independent modifications to the inspiral GW amplitude and phase,
while Sec.~\ref{Merger-RD-Sec} does the same for the merger and ringdown phase;
Sec.~\ref{ppE-framework} uses the results from the preceding sections to construct an particularly simple example of a ppE template bank;
Sec.~\ref{conclusions} concludes and points to future research. 
Appendix~\ref{bestiary} presents a brief summary of well-known alternative theories and their effect on
the GW observable, and Appendix~\ref{exotic-BHs} describes suggested
``exotic'' alternatives to BHs.

We follow here the conventions and notation of Misner, Thorne and Wheeler~\cite{Misner:1973cw}. 
In particular, Greek letters stand for spacetime time indices, while Latin letters stand for spatial indices only. 
Commas in index lists stand for partial differentiation, while semi-colons stand for covariant differentiation.
The Einstein summation convention is assumed unless otherwise specified. The metric signature is $(-,+,+,+)$
and we use geometric units where $G = c = 1$. In the Appendices, we reinstate the powers of G and c, as one
must distinguish the GR coupling constants from those in alternative theories of gravity.

\section{Anatomy of a Compact Object Binary Coalescence in GR}
\label{anatomy}

Let us now consider the quasi-circular binary coalescence of non-spinning 
BH-{\em like} compact objects---namely, events that exhibit inspiral, merger and 
ringdown phases akin to binary 
BH mergers in GR. Certain classes of ``exotic'' horizon-less compact objects, 
such as boson stars (see Appendix ~\ref{exotic-BHs}), 
may exhibit similar merger waveforms, and insofar as they are viable candidates
to explain compact objects in the universe that today are classified as BHs, they are also within
the scope of the ppE templates we will construct here. Neutron star mergers, for example, 
are {\em not} to be consider part of this family, as these are a separate
class of GW sources that would require a different ppE expansion (though ostensibly the
inspiral phase of a NS/NS merger ppE template would be the same as the BH/BH one, 
and if the merger regime for the former happens outside the detector's 
sensitivity window, as is expected to be the case with LIGO/GEO/Virgo, then
the binary BH ppE templates would suffice).

We divide the coalescence into three
stages: (i) the inspiral; (ii) the plunge and merger; (iii) the ringdown. In the first  stage,
the objects start widely separated and slowly spiral in via GW radiation-reaction. In the second
stage, the objects rapidly plunge and merge, roughly when the object's separation is somewhere 
around the location of the light-ring. In the third
stage, after a common apparent horizon has formed, the remnant rings down and settles to an equilibrium 
configuration. Note that this classification is somewhat {\emph{fuzzy}}, as studies of numerical 
simulation results have shown
that a sharp transition does not exist between them in GR~\cite{Buonanno:2006ui,Sperhake:2007gu,
Berti:2007fi}.
In the remainder of this section, we concentrate on GR coalescences and leave discussions about possible
modifications to the next section. 

In describing the waveform, we shall concentrate on the Fourier transform, ${\tilde{h}}(f)$, of the GW response function, $h(t)$, as the former is more directly applicable to GW data analysis (see eg.~\cite{Finn:1992wt,Jaranowski:2007pe} 
for a data analysis review) .
The response function of an interferometer, $h(t)$, is a time-series defined via the projection 
$h(t) \equiv F_{+} h_{+} + F_{\times} h_{\times}$, where $F_{+,\times}$ are {\emph{beam-pattern functions}} 
that describe the intrinsic detector response to a GW, while $h_{+,\times}$ are the projections of the GW metric perturbation 
$h_{ij}$ onto a plus/cross polarization basis. 
We will split the Fourier transform of the response function $\tilde{h}(f)$ into three components,
$\tilde{h}_{\I}(f)$,$\tilde{h}_{\M}(f)$ and $\tilde{h}_{\RD}(f)$, corresponding
to the inspiral, merger and ringdown phases respectively.
We will further decompose each of these complex functions into their real amplitude
and phase components:
\be
\tilde{h}_{\I,\M,\RD}(f) = A_{\I,\M,\RD}(f) \; e^{i \Psi_{\I,\M,\RD}(f)},
\label{amp-n-phase}
\ee
all of which will be modified in the ppE framework.

The inspiral phase can be modeled very well by the {\emph{restricted}} PN approximation (see eg.~\cite{Yunes:2009yz}). 
In this approximation, the amplitude of the time-series is assumed to vary slowly relative to the phase, allowing
its Fourier transform to be calculated via the 
stationary-phase approximation~\cite{Bender,Droz:1999qx,Yunes:2009yz}. Improving on the restricted
approximation leads to sub-leading amplitude corrections that introduce higher harmonics in the Fourier transform, 
though we shall not consider that here~\cite{Droz:1999qx,Arun:2007hu}. However, we should note that if considering
extreme-mass ratio inspirals (EMRIs) or eccentric orbits, then higher harmonics will play a significant role and should be accounted 
for (see eg.~\cite{Yunes:2009yz}).

In GR, within the restricted PN approximation the Fourier transform of the response function 
can be written as 
\ba
A_{\I}(f) &=& a_{\I} f ^{-7/6},
\\
\Psi_{\I}(f) &=& 2 \pi f t_{0} + \phi_{0} + \sum_{k=0} \psi_{k}^{\PN} u^{(k-5)/3},
\label{inspiral}
\ea
where we recall that $u \equiv \pi {\cal{M}} f$ is the reduced frequency parameter, 
${\cal{M}}=M \eta^{3/5}$ is the chirp mass, $\eta = m_{1} m_{2}/M^{2}$ is the symmetric mass
ratio, $M = m_{1} + m_{2}$ is the total mass and $f$ is twice the Keplerian orbital frequency.
The coefficient $a_{I}$ depends on the chirp mass, the luminosity distance $D_{L}$ 
and the orientation of the binary relative to the detector (parameterized through the inclination
$\iota$ and polarization angle $\theta$) via\footnote{When modeling the Fourier transform of the 
response function for LISA, one must multiply $a_{\I}$ by a factor of $\sqrt{3}/2$ to account for the geometry of the 
experiment. See eg.~Eq.~$(4.18)$ and $(5.6)$ of~\cite{Yunes:2009yz}. Moreover, the beam pattern functions
further depend on the motion of the detector around the Sun, which can be modeled as in eg.~\cite{Cutler:1997ta,Finn:1992xs}}:
\be
a_{\I} = - \left( \frac{5}{24} \right)^{1/2} \pi^{-2/3} \frac{{\cal{M}}^{5/6}}{D_{L}} Q(\iota,\theta),
\label{Q-disc}
\ee
where $Q(\iota,\theta) = F_{+} \cos{2 \theta} (1 + \cos^{2}{\iota})/2 - F_{\times} \sin{2\theta} \cos{\iota}$ 
is a function of the beam-pattern functions $F_{+,\times}$, the inclination $\iota$ and the polarization angle 
$\theta$.\footnote{Note that the function $Q$ as defined here differs from that defined in~\cite{Yunes:2009yz} 
by a factor of $4$.} The
quantities $t_{0}$ and $\phi_{0}$ are constants, related to the time and phase of coalescence, 
while $\psi_{k}^{\PN}$ are constant PN phase parameters that depend on the mass ratio. To $1.5$ PN order, these coefficients are given by
\ba
\psi_{0}^{\PN} &=& \frac{3}{128},
\nonumber \\
\psi_{1}^{\PN} &=& 0,
\nonumber \\
\psi_{2}^{\PN} &=& \frac{3}{128} \left[\frac{20}{9} \left(\frac{743}{336} + \frac{11}{4} \eta \right) \eta^{-2/5}\right],
\nonumber \\
\psi_{3}^{\PN} &=& \frac{3}{128} \left(-16 \pi \eta^{-3/5} \right),
\label{PN-phase-par}
\ea
and expressions to higher order are also known~\cite{Blanchet:2001ax}.

The merger phase cannot be modeled analytically by any controlled perturbation scheme. We shall thus treat this phase as
a transition region that interpolates between the inspiral and the ringdown. During this phase, numerical simulations have shown 
that the amplitude and the phase can be fit by~\cite{Buonanno:2006ui,Ajith:2007kx}
\ba
A_{\M}(f) &=& a_{\I} f_{\IM}^{-1/2} f ^{-2/3},
\\
\Psi_{\M}(f) &=& \bar{\phi}_{0} + 2 \pi f \bar{t}_{0}
\ea
where $f_{\IM}$ is the frequency of transition between inspiral and merger and where 
$\bar{\phi}_{0}$ and $\bar{t}_{0}$ are new constants, and can be related to ${\phi}_{0}$ and ${t}_{0}$
by demanding continuity of the waveform. The frequency dependence of the amplitude in the merger phase,
$f^{-2/3}$,is not universal,
but specific to the test system (equal mass, non-spinning) studied 
here\footnote{Though note that in~\cite{Baumgarte:2006en} it was suggested
that the exponent might be closer to $-5/6$.}. 

The ringdown phase can be modeled very well through BH perturbation theory techniques 
(see eg.~\cite{Berti:2009kk} for a recent review). 
In this approximation, the GW response function can be written as a sum of
 exponentially damped sinusoids, 
the Fourier transform of each being a 
Lorentzian function. Furthermore, for equal mass mergers,
a single (least damped) QNR mode dominates the waveform,
and so to good approximation we can write~\cite{Ajith:2007kx}
\ba
A_{\RD}(f) &=& a_{\I} f_{\IM}^{-1/2} f_{\MRD}^{-2/3} \left[ 1 + 4 \left(\frac{f - f_{\MRD}}{\sigma} \right)^{2} \right]^{-1},
\label{RD-A}
\\
\Psi_{\RD}(f) &=& \bar{\phi}_{0} + 2 \pi f \bar{t}_{0},
\label{RD-P}
\ea
where $\sigma$ is the width of the Lorentzian that essentially describes the mean duration of the ringdown phase (the damping time).
Notice that during the plunge and the ringdown the phase remains the same, modulo non-linear effects, which conveniently guarantees
continuity of the phase across this boundary. 

The phenomenological parameters $(f_{\IM},f_{\MRD},\sigma)$ can be fit to numerical simulations to find~\cite{Ajith:2007kx}
\ba
f_{\IM} &\approx& \frac{0.29 \eta^{2} + 0.045 \eta + 0.096}{\pi \; M} \to  \frac{1}{8 \pi M},
\nonumber \\
f_{\MRD} &\approx& \frac{0.054 \eta^{2} + 0.090 \eta + 0.19}{\pi \; M} \to  \frac{1}{4 \pi M},
\nonumber \\
\sigma &\approx& \frac{0.51 \eta^{2} + 0.077 \eta + 0.022}{\pi \; M}  \to \frac{1}{14 \pi M},
\ea
where the arrow stands for the approximate value for equal mass binaries. The quantities $(f_{\IM},f_{\MRD})$
denote the boundaries of the inspiral-merger region and merger-ringdown region, at which the piece-wise function
has been constructed to be continuous. Notice that the merger and ringdown phases are both rather short compared
to the inspiral phase, lasting approximately $(8 \pi M)^{-1}$ Hz and $(14 \pi M)^{-1}$ Hz respectively. Due to the short
duration of these phases, they can be modeled via the simple linear relation in the phase shown above, and in fact, 
as the merger phase is fairly constant, it could be absorbed in the overall amplitude.

The parameterization presented here is purposely similar to that proposed in~\cite{Ajith:2007kx}. In that study, however, 
the phase was modeled in all stages of coalescence by the same function, namely Eq.~\eqref{inspiral}, where the
parameters $\psi_{k}^{\PN}$ are not fixed to the PN value, but instead fitted to numerical relativity waveforms 
(see eg.~Table $2$ in~\cite{Ajith:2007kx}). 
If one prefers, one could model the GW phase in the merger and ringdown with the fitted parameters presented
in Table $2$ of~\cite{Ajith:2007kx}. 
Alternatively, one could also follow the prescription of~\cite{Berti:2005ys} and construct the Fourier transform of the 
response function for an infinite sum of modes, absorbing the phase into a complex amplitude and paying more careful attention 
to the response function, but we leave such details for future work.  

\section{Model-Independent Modifications to the Inspiral Phase}
\label{Inspiral-Sec}

We shall now consider {\emph{leading-order}} deviations from GR in the inspiral phase. 
In the restricted PN and stationary phase approximations,
the Fourier transform of the response function can be written as (see eg.~Eq.~$(4.5)$ in \cite{Yunes:2009yz})
\be
\label{spa-FT}
\tilde{h}_{\I}(f) \equiv A_{\I}(f) e^{i \Psi_{\I}(f)}  = \frac{{\cal{A}}(t_{0})}{2} \; \frac{1}{\sqrt{2 \dot{F}}} \;  e^{i \Psi(f/2)},
\ee
where ${\cal{A}}(t_{0})$ is the time-domain amplitude evaluated at the stationary point,
defined via $\dot\Psi(t_{0}) = \pi f$, and overhead dots stand for partial differentiation with respect to time. 
$F$ is the orbital frequency, and $\dot{F}$ is its rate of change
evaluated at the stationary point $F(t_{0}) = f/2$; recall that 
$f$ is the GW frequency. The quantity $\Psi(F)$ is the GW 
phase, which is given by (see eg.~Eq.~$(4.8)$ in~\cite{Yunes:2009yz})
\be
\label{SPA-phase}
\Psi_{\I}(f) \equiv - \frac{\pi}{4} - 2 \pi \int^{f/2} dF'   \left(2 - \frac{f}{F'}\right) \tau(F'),
\ee
where $\tau(F) \equiv F/\dot{F}$.
The structure of Eq.~\eqref{spa-FT} is generic for any Fourier transform 
with a generalized Fourier integral that contains a stationary
point~\cite{Bender}.

The orbital frequency and its rate of change completely determine the Fourier transform of the response function in the
stationary phase approximation. The rate of change of the orbital frequency is generically computed through the product
rule $\dot{F} = \dot{E} \left(dE/dF\right)^{-1}$, where 
$E(F)$ is the orbital binding energy. 
The quantity $\dot{E}$ is calculated by invoking the so-called {\emph{balance law}}. 
This law states that the amount of energy carried away in GWs is equal
to the amount of orbital binding energy lost by the binary system, {\emph{i.e.~}}$\dot{E} = -{\cal{L}}_{\GW}$, where ${\cal{L}}_{\GW}$ is the
GW luminosity. The luminosity can be computed perturbatively in terms of a multipolar 
decomposition of the effective energy-momentum tensor of the system~\cite{Thorne:1980rm}, 
which to leading order yields $\dot{E} \sim |d^{3}I_{ij}/dt^{3}|^{2}$, where $I_{ij}$ is the reduced quadrupole 
moment tensor of the binary.

We will consider GR modifications of the Fourier transform of the response function 
during the inspiral arising either from changes to the 
conserved binary binding energy (the Hamiltonian, see eg.~\cite{lrr-2006-4} for a more precise definition),
or changes to the energy balance law. The former are essentially 
modifications to the {\emph{conservative}} dynamics, while the latter modifies the {\emph{dissipative}} dynamics. 
We shall tackle each of these corrections next.

\subsection{Modifications to the Binary Hamiltonian}
\label{mod-bin-ham}

The Hamiltonian for a binary system, or equivalently, the binary's center of mass binding energy, 
can usually be expressed as a linear combination of kinetic energy $T$ and potential energy $V$:
$E = T(p) + V(q,p)$, where $q$ and $p$ stand for generalized coordinates and conjugate momenta.
For a binary system, one can extract the Newtonian contribution and rewrite such a generic Hamiltonian as
\be
\frac{E}{\mu} = \frac{v^{2}}{2} - \frac{M}{r_{12}} + V(v,r_{12}),
\label{H-par}
\ee
where $v = |v^{i}_{1} - v^{i}_{2}|$ is the relative velocity, $\mu = \eta M$ is the reduced mass, 
$r_{12}$ is the orbital separation, and $M$ and $\eta$ 
are as before the total mass and symmetric mass ratio.
The quantity $V(v,r_{12})$ is some additional contribution to the Hamiltonian that can be due either to relativistic effects, or
to non-GR contributions, for example from additional scalar or dipolar degrees of freedom.
Of course, not all functions $V$ are allowed, since binary pulsar observations have already verified GR in a 
quasi-weak field,
which translates into the requirement $V \ll M/r_{12} \ll 1$ when $v \ll 1$. 

From such a Hamiltonian, one can derive a modified version of Kepler's third law (or equivalently Newton's second law),
namely
\be
\label{Kepler}
\omega ^{2} = \frac{M}{r_{12}^{3}} + \frac{1}{r_{12}} \Omega(v,r_{12}),
\ee
where $\omega$ is the binary's orbital angular frequency and $\Omega \equiv dV/dr_{12}$. 
We see then that modifications to the Hamiltonian and to Kepler's law
are not independent, as expected.

Kepler's law must be inverted to obtain the separation as a function of angular frequency, which can then be inserted into the 
Hamiltonian to obtain $dE/dF$. Although the unspecified form of $\Omega$ does not allow 
an explicit inversion, we can always write the inversion {\emph{implicitly}} as
\be
r_{12} = M^{1/3} \omega^{-2/3} + \chi(\omega),
\label{inv-r12}
\ee
where $\chi$ is some function of $\omega$ determined from the inversion of Eq.~\eqref{Kepler}. 
As for $(V,\Omega)$, the quantity $\chi$ must also satisfy the condition $\chi \ll M^{1/3} \omega^{-2/3}$ 
in the weak field.

One possible parameterization of $\chi$ that satisfies the above condition is
\be
r_{12} = M x^{-1} \left[1 + \lambda_{1} \; x^{p}\right],
\label{r12-par}
\ee
where $x \equiv (M \omega)^{2/3}$ is a PN parameter, while $\lambda_{1}$ and $p$ are dimensionless, real numbers. 
The parameters $\lambda_{1}$ and $p$ determine roughly where the non-Newtonian correction 
to Kepler's law begins to become important, and they must be real since $\omega$ is real.
Furthermore, the condition that $\chi \ll M^{1/3} \omega^{-2/3}$ when $x \ll 1$ implies that $p > 0$ in the inspiral phase.
The choice of parameterization of $\chi$ in Eq.~\eqref{r12-par} is equivalent to choosing a parameterization for the potential 
$V$ in Eq.~\eqref{H-par}; with our choice one finds that
\be
V = - \frac{3 \lambda_{1}}{1 + p} x^{1+ p},
\label{V-par}
\ee
to first order in $x$.

We have here chosen a parameterization for $\chi$ that allows an inversion analytically in terms of known functions.  
A more general parameterization would be to add $N$ terms with exponents $d_{i,N} > d_{i,N-1} > \ldots > p > 0$. 
However,  the higher the exponent, the smaller its contribution, since the inspiral phase 
must terminate outside the light-ring where in GR $x < 0.4$ (for simplicity, we here work with one exponent only). 
In fact, in GR one finds that perturbatively 
\ba
r_{12} &=& M x^{-1} \left\{
1 +  
\frac{1}{3} x \left(\eta - 3\right) 
\right.
\nonumber\\
&+&  \left.
x^{2} \left(\frac{19}{4} \eta + \frac{1}{9} \eta^2\right) + 
x^{3} \left[- \frac{24257}{2520} \eta - \frac{37}{12} \eta^2 
\right. \right. 
\nonumber \\
&+& \left. \left.
\frac{2}{81} \eta^3 + \frac{41}{192} \eta \pi^2 
+ \frac{22}{3} \eta \ln\left(\frac{x_{0}}{x} \right) \right]  
\right\},
\ea
when $x \ll 1$ from Eq.~$(190)$ of~\cite{lrr-2006-4}, where here 
$x_{0}$ is a gauge parameter associated with harmonic coordinates. 
From the GR expression, we also infer that the constant $\lambda_{1}$ could depend on 
the system parameters, such as the mass ratio. Also, $\lambda_{1}$ must depend on some non-GR coupling 
intrinsic to the alternative theory, such that in the limit as this parameter vanishes, one recovers the GR prediction. 

The binary's center-of-mass energy can now be expressed entirely in terms of the orbital frequency. 
Requiring the circular orbit condition $v = r_{12} \omega$, we find
\be
\frac{E}{\mu} =
\frac{1}{2}  x \left[1 + \lambda_{1} x^{p}\right]^{2}
- x \left[1 + \lambda_{1} x^{p}\right]^{-1}
+ V(x).
\ee
In principle, we should not expand in $x \ll 1$, since we are interested in corrections that could arise in the late 
inspiral (though we know that even close to merger and in GR $x < 0.4$). To simplify the following analysis,
we shall perform this expansion regardless, as it only serves to motivate the final ppE parameterizations, where $x$
can take large values in the late inspiral.
Performing the $x \ll 1$ expansion, we find  
\be
\frac{E}{\mu} =  -\frac{1}{2}  x \left[1 + \left(\frac{6}{1 + p} - 4\right) \lambda_{1} x^{p}\right],
\ee
and differentiating with respect to $F = \omega/(2 \pi) $ we obtain
\be
\frac{dE}{dF} = - \frac{2 \pi}{3} \eta M^{2} x^{-1/2} \left[ 1 + 2 \lambda_{1} \left(1 - 2 p\right) x^{p} \right]
\ee

The above considerations provide motivation to extrapolate a model-independent, non-GR parameterization for $dE/dF$: 
\be
\frac{dE}{dF} = - \frac{2 \pi}{3} {\cal{M}}^{2} u^{-1/3} \left[ 1 + \lambda_{2} \; u^{q} \right].
\ee
where we recall that $u \equiv \pi {\cal{M}} f = 2 \pi {\cal{M}} F = x^{3/2} \eta^{3/5}$ for circular orbits.  
We have simplified the above parameterization by introducing $\lambda_{2} \equiv 2 \lambda_{1} (1 - 2 p)$ and $q \equiv 2 p/3$. 
Such an expression resembles the PN expansion of GR, which one can derive from Eq.~$(194)$ of~\cite{lrr-2006-4}:
\ba
\frac{dE}{dF} &=& - \frac{2 \pi}{3} {\cal{M}}^{2} u^{-1/3} \left\{
1 +
u^{2/3} \eta^{-2/5} \left(-\frac{3}{2} - \frac{1}{6} \eta\right) 
\right. 
\nonumber \\
&+& \left. 
u^{4/3} \eta^{-4/5} \left[- \frac{81}{8} + \frac{57}{8} \eta - \frac{1}{8} \eta^2\right] + 
u^{2} \eta^{-6/5}  
\right.  
\\ \nonumber 
&\times& \left.
\left[ -\frac{675}{16} +
\frac{34445}{144} \eta - \frac{205}{24} \eta \pi^2 - \frac{155}{24} \eta^2 - \frac{35}{1296} \eta^3\right].
\right\}
\ea
Similarly, the parameter $\lambda_{2}$ could depend on $\eta$ but it must also be proportional
to a positive power of the alternative theory couplings, such that when these vanish, one recovers the PN 
expansion of the GR expectation.

\subsection{Modifications to the Energy Balance Law}
\label{mod-E-law}

The energy flux at infinity is usually calculated in GR by assuming that the only source of radiation are GWs, so that 
$\dot{E} = - {\cal{L}}_{\GW}$, where we recall that ${\cal{L}}_{\GW}$ is the GW luminosity. In alternative theories of gravity, 
however, there could be additional sources of radiation from other propagating degrees of freedom. The balance law
then becomes $\dot{E} = - {\cal{L}}_{\GW} - {\cal{L}}_{\rm other}$. 

The calculation of the GW luminosity is generically performed in two steps. First, one employs perturbation theory to determine
the effective (Isaacson) GW stress-energy tensor~\cite{Isaacson:1968ra,Isaacson:1968gw}, the time-time component of which 
provides an expression for ${\cal{L}}_{\GW}$ in terms of time derivatives of the metric perturbation $h_{ij}$. In GR, this expression is
schematically ${\cal{L}}_{\GW} \sim \dot{h}_{ij} \dot{h}^{ij}$. Second, one performs a multipolar decomposition of the metric into mass
and current multipoles~\cite{Thorne:1980rm}, which are given by integrals over the sources and for point sources are simply proportional
to their trajectories and velocities. In GR, one then schematically finds $h_{ij} \sim \ddot{I}_{ij}/R$, where $I_{ij}$ is
the reduced quadrupole moment and $R$ is the distance to the observer.
Combining both steps into one single procedure, one finds that the GW luminosity in GR is proportional to third time-derivatives of the 
reduced quadrupole moment ${\cal{L}}_{GW} \propto \dddot{I}_{ij} \dddot{I}^{ij}$. 

In general, one can decompose the GW luminosity in a superposition of mass and current multipole moments~\cite{Thorne:1980rm} 
\be
{\cal{L}}_{GW} \propto \kappa_{M} \dot{\cal{M}}^{2} + \kappa_{D} \ddot{D}_{i} \ddot{D}^{i} + \kappa_{Q}  \dddot{I}_{ij} \dddot{I}^{ij}
+ \ldots,
\label{lum-gen}
\ee
where ${\cal{M}}$ is the monopole of the system, $D$ is its dipole moment and $\kappa_{M,D,Q}$ are proportionality constant for the 
monopole, dipole and quadrupole contributions respectively. 
The quantity $\kappa_{M}$ has the effect of re-normalizing Newton's gravitational constant $G$, and thus, can be bundled together 
with $\kappa_{Q}$.
On the other hand, the quantity $\kappa_{D}$ is related to the self-binding energy of the gravitating bodies, and it produces an independent
contribution to the GW luminosity.  
In GR, these two contributions (monopole and dipole) identically vanish due to center-of-mass and 
linear-momentum conservation. 
Evaluating Eq.~\eqref{lum-gen} with standard definitions of multipole moments for a binary system~\cite{Thorne:1980rm}, one finds to leading order
\be
\dot{E} = - \frac{\mu^{2} M^{2}}{r_{12}^{4}} \left[ \frac{8}{15} \left(\kappa_{1} v^{2} - \kappa_{2} \dot{r}^{2} \right) + \frac{1}{3} \kappa_{D} {\cal{G}}^{2} \right] - {\cal{L}}_{\rm other},
\ee
where $\kappa_{1,2}$ are the so-called Peters-Mathews parameters, whose numerical value depends on the alternative theory under 
consideration, while $\kappa_{D}$ is a dipole self-gravitational energy contribution, with ${\cal{G}}$ the difference in self-gravitational 
binding energy per unit mass (see eg.~Eq.$(10.84)$ in~\cite{Will:1993ns}). 
For example, in GR $(\kappa_{1}, \kappa_{2}, \kappa_{D}) = (12,11,0)$, while in Brans-Dicke theory without a cosmological constant 
$[\kappa_{1}, \kappa_{2}, \kappa_{D}] = [12 - 5 \; (2 + \omega_{\BD})^{-1},11 - 11.25 \; (2 + \omega_{\BD})^{-1},2 \; (2 + \omega_{\BD})^{-1}]$~\cite{Will:1993ns}, where $\omega_{\BD}$ is the BD coupling parameter.

Using the modified Kepler's law from the previous subsection, we can re-express the power as a function of frequency only, namely
\ba
\dot{E} &=& -\frac{32}{5} \eta^{2} x^{5} 
\left[ \frac{\kappa_{1}}{12} \left(1 - 2 \lambda_{1} x^{p} \right) 
\right. 
\nonumber \\
&+& \left.
\frac{5}{96} \kappa_{D} {\cal{G}}^{2} x^{-1} \left(1 - 4 \lambda_{1} x^{p}\right)\right] 
- {\cal{L}}_{\rm other},
\label{mod-P}
\ea
where we have set $\dot{r} = 0$, corresponding to a circular orbit. 
The dipolar term dominates over GR's quadrupolar emission during the early inspiral, which
explains why Brans-Dicke theory, whose dominant signature is precisely dipolar radiation (see Appendix~\ref{ST-app}), 
has been constrained so well by binary pulsar observations. 
On the other hand, the modification to Kepler's law becomes important as the binary approaches the end of the inspiral phase, 
as expected since it is meant to model dynamical strong-field corrections to GR.  

The quantity ${\cal{L}}_{\rm other}$ serves to parameterize both additional contributions from scalar or vectorial degrees of freedom, 
as well as high-order curvature corrections to the GW effective stress-energy tensor. 
In the case of the former, one could for example have ${\cal{L}}_{\rm other} = {\cal{L}}_{\phi} \propto \dot{\phi}^{2}$, 
where $\phi$ is some scalar field that is activated when the binary approaches the merger phase. 
In fact, such is the case in dynamical CS modified gravity (see Appendix~\ref{CS-app}), where the CS field becomes strongly sourced
during the late inspiral, plunge and merger, as the Riemann curvature grows. 
Here, one could for example have 
${\cal{L}}_{\rm other} = {\cal{L}}_{R^{\star}R} \propto |\dddot{I}_{ij} \cdot (\nabla_{k} \dddot{I}_{l m})^{2}|^{2}$,
an interaction that is absent in GR.

\subsection{Modified GW Phase}

We can now combine the results from the previous two sections to find $\dot{F}$:
\be
\dot{F} = \frac{48}{5 \pi {\cal{M}}^{2}} u^{11/3} 
\left[ \bar{\kappa}_{1} + \lambda_{3}  u^{q} 
+ \bar{\kappa}_{D} 
u^{-2/3} 
+ \widetilde{\cal{L}}_{\rm other} u^{-10/3} 
\right],
\label{Fdot}
\ee
where we have rescaled $\bar{\kappa}_{1} \equiv \frac{\kappa_{1}}{12}$, 
$\bar{\kappa}_{D} \equiv \frac{5}{96} \kappa_{D} {\cal{G}}^{2} \eta^{2/5}$,
$\widetilde{\cal{L}}_{\rm other} \equiv \frac{5}{32} {\cal{L}}_{\rm other}$, and
where the new constant parameter $\lambda_{3}$ can be related to 
$(\kappa_{1},\lambda_{1},p,\eta)$, though the expression is not particularly illuminating.
We have also here dropped second-order terms, such as the influence of the modified 
Keplerian relation on dipolar emission. 
The parameter $\bar{\kappa}_{1}$ has the effect of re-normalizing the value of $G$ or $c$, which would 
not be directly observable in a GW detection, as it would be directly reabsorbed by physical quantities, like the chirp mass 
(unless one possesses an electromagnetic counterpart).
The parameter $\bar{\kappa}_{D}$ parameterizes dipolar deviations and depends on the sensitivities
of the bodies. 
The parameters $\bar{k}_{1}$ and $q$ parameterize deviations from Kepler's law, while 
$\bar{{\cal{L}}}_{\rm other}$ describes possible additional sinks of energy.

We can now integrate Eq.~\eqref{SPA-phase} to obtain the modified GW phase:
\ba
\Psi_{\I}(f) &=& - \frac{\pi}{4} - \phi_{c} + 2 \pi f t_{c} 
+ \frac{3}{128 \pi} u^{-5/3}  
\nonumber \\
&\times& 
\left[ \bar{\kappa}_{1}  - \lambda_{4} u^{q} - {\bar{\bar{\kappa}}}_{D} u^{-2/3}  
 \right]
\label{insp-GW-phase-gen}
\\ \nonumber   
&+&
\frac{10 \pi {\cal{M}}}{48} \int^{f/2} \left(2 - \frac{f}{F'}\right) (u')^{-18/3} 
{\widetilde{{\cal{L}}}}_{\rm other} 
dF',
\ea
where $\lambda_{4}$ and ${\bar{\bar{\kappa}}}_{D}$ are new parameters that again can be related to the 
old ones, though these relations are unilluminating.
As is customary, we have here linearized in the GR corrections to be able to perform the integral analytically, 
an approximation that is valid in the inspiral, although not necessarily so in the plunge and merger. 
Note that the term proportional to ${\bar{\bar{\kappa}}}_{D}$ mimics the Brans-Dicke correction to the GW phase, which is known to scale as $u^{-7/3}$
(see discussion around Eq.~\eqref{betaBD}), while $q = -1$ mimics the effects of a massive graviton on the propagation of GWs, 
which is known to scale as $u^{-1}$ (see discussion around Eq.~\eqref{betaMG}). 
 
\subsection{Modified GW Amplitude}

The GW amplitude depends both on the amplitude of the time-domain waveform evaluated at the stationary point and the rate of change 
of the frequency $\dot{F}$. The latter has already been parameterized in Eq.~\eqref{Fdot}. The former, however, requires
further study. In GR, the quadrupole formula requires that 
\be
h_{ij} \propto \frac{1}{D_{L}} \ddot{I}_{ij},
\label{quadrupole}
\ee
which leads to the following waveform amplitude to leading order for a binary system 
\be
{\cal{A}}(t) = -\frac{4 \mu}{D_{L}} \omega^{2} r_{12}^{2} Q(\iota,\beta),
\ee
where $Q(\iota,\beta)$ is as defined after Eq.~\eqref{Q-disc}.
Evaluating this amplitude at the stationary point and using
the modified Keplerian relations of Eq.~\eqref{r12-par}, we find
\be
{\cal{A}}(t_{0}) = - \frac{4 {\cal{M}}}{D_{L}} Q(\iota,\beta) u^{2/3} \left[ 1 + 2 \lambda_{1} \eta^{-2 p/5} u^{q} \right].
\ee
Combining this expression with Eq.~\eqref{Fdot}, we find that $A_{\I} \equiv |\tilde{h}_{\I}|$ is equal to  
\ba
A_{\I}(f) &=& -\left(\frac{5}{24}\right)^{1/2} \pi^{-2/3} \frac{{\cal{M}}^{-5/6}}{D_{L}} Q(\iota,\beta) f^{-7/6} \left( \bar{\kappa}_{1} 
\right.
\nonumber \\
&+& \left.
\tilde{k}_{\I} u^{q} 
-\frac{\kappa_{D}}{2}  u^{-10/3} - \frac{1}{2} \widetilde{\cal{L}}_{\rm other} u^{-10/3} \right)
\label{insp-GW-amp-gen},
\ea
where $\tilde{k}_{\I}$ is another constant and we now kept only the leading-order corrections. 
Such an expression agrees identically with the GR expectation when $\bar{\kappa}_{1} = 1$ 
and all other modified parameters vanish. 

The above results hinge on the assumption that the quadrupole formula in Eq.~\eqref{quadrupole} holds to leading order 
in the GR modification. In well-known alternative theories of gravity, including scalar-tensor ones, Eq.~\eqref{quadrupole} is indeed valid
to leading order, acquiring subdominant corrections. These corrections can be modeled by the undetermined function $\widetilde{\cal{L}}_{\rm other}$,
in the same fashion as in Sec.~\ref{mod-E-law}. Thus, the generalization of Eq.~\eqref{quadrupole} leads to redundancies in the parameterization
that we are after. In Sec.~\ref{ppE-framework} we shall use the insight gained from the analysis of this
section to infer a generalized parameterization for a model-independent modification of the inspiral gravitational waveform.

\section{Model-Independent Modifications to the Merger and Ringdown Phase}
\label{Merger-RD-Sec}

\subsection{Merger Phase}

Let us now consider {\emph{leading-order}} deviations from GR first during the merger and then the ringdown phase.
The merger phase is the least understood regime of a binary coalescence, because no
controlled approximation scheme can be employed to solve the field equations. 
In GR, numerical simulations of the merger of two, non-spinning, equal-mass BHs has produced the phenomenological fit 
\be
\tilde{h}_{\M}(f) \propto u^{-2/3} e^{i \psi_{M}(f)},
\ee
where the phase is a linear function $\Psi_{M}(f) = \delta + \epsilon u$. 
Comparing to the results of Sec.~\ref{anatomy} we find that $\delta = \bar{\phi}_{0}$ 
and $\epsilon = 2 \bar{t}_{0}  {\cal{M}}^{-1}$.
Linearity in the phase can be assumed due to the short duration of the merger phase (usually less than
2 GW cycles for BHs). This functional form is an approximate fit and found only {\emph{a posteriori}},
{\emph{i.e.~}}after a full numerical solution has been obtained. 

Lack of formally derivable analytic approximations of the merger waveform is not 
exclusive to GR. Furthermore, due to the dearth of numerical studies of this
regime in alternative theories (but see~\cite{Salgado:2008xh} for promising plans to explore
mergers in scalar-tensor theories), little guidance can be found on how 
the merger regime may differ from GR. 
How do we then parameterize model-independent corrections of the merger waveform when we possess no analytic or 
numerical guidance?  
For lack of a better prescription, we will here assume that the waveform produced during the
merger can adequately be described as an {\emph{interpolation}} between the inspiral 
and the ringdown waveforms. This is a good approximation to GR waveforms, and hence 
is a rather conservative approach to a ppE extrapolation.
Specifically then, we propose
\be
\tilde{h}_{\M}(f) = \sum_{n=0}^{\infty} \gamma_{\M,n} u^{c_{\M,n}} e^{i(\delta + \epsilon u) },
\label{merger-wave}
\ee
where $\gamma_{\M,n}$ and $c_{\M,n}$ are an infinite set of undetermined parameters. We shall here 
keep only one term in the polynomial, setting $\gamma_{\M,0} = \gamma$, $c_{\M,0} = c$ and all other amplitude parameters 
to zero, while we shall leave $\epsilon$ undetermined in the phase. The parameters $(\gamma,\delta)$ are set be requiring 
continuity between the inspiral-merger interface at $u=u_{\IM}$: 
$\gamma = A_{\I}(f_{\IM}) u^{-c}_{\IM}$ and $\delta = \Psi_{\I}(f_{\IM}) - \epsilon u_{\IM}$.
The choice of a single term in the power-series representation of the Fourier transform of the merger response function forces us to
consider coalescences with prompt mergers. Delayed mergers can occur, for example,
with exotic compact objects such as boson stars~\cite{Palenzuela:2007dm}, in 
a process akin to delayed-collapse scenarios in NS mergers.
Of course, additional terms could be added to the series of Eq.~\eqref{merger-wave}
to account for such scenarios, however for simplicity here we will restrict attention
to prompt mergers.

\subsection{Ringdown Phase} 
Let us now consider the {\emph{leading-order}} modification to GWs that could be produced during 
the post-merger phase due to alternative theory corrections. 
GWs emitted in this stage are produced by oscillations in the remnant spacetime
as it settles down to its final equilibrium configuration. Simulations show
that in GR the ringdown very quickly becomes linear, and we will assume
the same here for the ppE expansion. In GR, the metric perturbation
in general is a sum of quasi-normal ringdown (QNR) modes, which are exponentially
damped sinusoids, and so-called tail terms, which are constant phase solutions
that decay via a temporal power law. The QNR modes dominate the early-time
post-merger waveform, which is why we label this phase the ringdown phase.

We can write the ringdown time-domain response function as
\be
h_{\RD}(t) = A_{\RD}(t) \cos[\Psi_{\RD}(t)],
\label{RD-wave}
\ee
where the phase is purely real and the amplitude is a decaying function
of time. Due to the assumed linearity, 
we can then decompose the amplitude and phase into a sum of harmonics
$h_{\RD} = \sum_{n,\ell,m} A^{n \ell m}_{\RD}(t) \cos[ \Psi^{n \ell m}_{\RD}(t)]$, 
where $(\ell,m)$ labels a spheroidal harmonic mode, and $n$ is an integer
describing the {\em overtone} of the given mode. For the QNR component
of each mode, the amplitude is a decaying exponential
\be
{}^{\rm QNR}A_{\RD}^{n\ell m}(t) =\theta(t) A_{0}^{n \ell m} e^{-t/\tau_{n \ell m}},
\label{RD-amp-1}
\ee
where $\tau_{n \ell m}$ is the decay constant of the mode, 
$\theta(\cdot)$ is the Heaviside function that guarantees the waveform is well
defined at negative infinite time, and $A_{0}^{n \ell m}$ is some constant amplitude determined 
by continuity  with the merger phase, which as such must depend on the beam-pattern functions. 
The phase of each QNR mode is
\be
{}^{\rm QNR}\Psi_{\RD}(t) = 2 \pi f_{\ell m n} t,
\label{RD-phase}
\ee
where $f_{\ell m n}$ is the real quasinormal frequency of the $(n,\ell,m)$ mode. 

In GR, the Fourier transform of the QNR component of the ringdown waveform is thus
a sum of Lorentzian-like functions:
\ba
{}^{\rm QNR}\tilde{h}_{\RD}(f) &=& \sum_{n \ell m} \frac{A^{n \ell m}_0}{2} \left(b_{+} + b_{-}\right),
\nonumber \\
b_{+,-} &=& \frac{\tau_{n \ell m}}{1 + 4 \pi^{2} \tau_{n \ell m}^{2} \left(f \pm f_{n \ell m}\right)^{2}}.
\label{GR-RD}
\ea
This result can be obtained by rewriting the cosine in Eq.~\eqref{RD-wave} as a sum of complex exponentials
using Euler's formula, and then following the {\emph{doubling prescription}} of~\cite{Flanagan:1997sx}, where one essentially 
replaces $t/\tau_{\ell mn} \to |t|/\tau_{\ell m n}$ in the Fourier integral and then multiplies the result by a correction factor that accounts 
for the doubling~\cite{Berti:2005ys}.

Tail terms in the post-merger waveform are 
induced either by linear interactions between the GWs and the background, or non-linear behavior right after 
merger.
Linear tails have a long history in GR~\cite{Price:1971fb,Price:1972pw,Gundlach:1993tp} and they are controlled 
by the asymptotic form of an effective potential far from the source. Such tails depend on the harmonic one is 
looking at but generically for a Schwarzschild background they can be parameterized by 
\be
{}^{\tail}A^{n\ell m}_{\RD}(t) = A_{0}^{n \ell m} t^{-(2 \ell +3)},
\quad
{}^{\tail}\Psi^{n\ell m}_{\RD}(t) = 0.
\ee
Recently in~\cite{Okuzumi:2008ej} it was suggested that ``non-linear tails'' could be produced during merger.
Such tails are structurally similar to the linear ones, except that one replaces the power $2 \ell + 3 \to b$, where $b$ is some positive integer.
In this case, the Fourier transform is given by
\be
{}^{\tail}\tilde{h}_{\RD}(f) = \sum_{n \ell m} - i \pi A_{0}^{n \ell m} \frac{\left(- 2 \pi i f\right)^{b-1}}{\left(b - 1\right)!},
\label{GR-RD-tail}
\ee
for positive frequencies (the Fourier transform for a Schwarzschild linear tail can be easily inferred from the above equation).
In GR then, one expects the complete ringdown GW to be a superposition of QNR terms
and tail terms, leading to a Fourier transform that is the linear combination of Eq.~\eqref{GR-RD} and~\eqref{GR-RD-tail}.

What generic type of modifications could we expect from alternative theories of gravity? 
One possibility would be to modify
the power in the exponential damping part of the ringdown, {\emph{i.e.~}}in Eq.~\eqref{GR-RD} 
$\exp(-t/\tau_{n \ell m}) \to \exp (-t^a/\tau_{n \ell m}^{a})$, with a positive real number. 
However, $a\neq1$ does not seem to be very ``natural'', and would significantly
complicate the analysis. For this study we simply keep $a=1$.
Another possible modification is to correct the tail behavior of the ringdown waves. Linear tails, however, 
are predominantly controlled by the dominant $1/r$ behavior of an effective potential in the far-field field, which in turn
depends on the $1/r$ behavior of the background metric tensor. If we require that all stable post-merger metric tensors approach 
the Schwarzschild solution at late times in the far-field (by Solar System tests), it is unclear how much linear tails would
be modified by non-linear terms in the gravitational field. On the other hand, non-linear tails could be modified in alternative
theories of gravity, as suggested in~\cite{Ching:1994bd,Ching:1995tj}. Nonetheless, alternative theory corrections to either linear 
or non-linear tail terms can both be parameterized by some undetermined power $b$, leading to the Fourier transform shown in 
Eq.~\eqref{GR-RD-tail}.

For simplicity then, we will propose a modified ringdown waveform with a similar functional form
as the GR result given in Eq.~\eqref{GR-RD}, though allow the decay constants, frequencies
and powers to take on non-GR values.

\section{Parameterized post-Einsteinian Framework}
\label{ppE-framework}

In this section we collate the results from the previous sections to 
construct ppE extensions of the inspiral, merger and 
ringdown waveform segments of a binary merger. In each of these phases, we shall divide this task into the construction of a ppE phase 
and a ppE amplitude, such that the waveform is given by: 
\be
\tilde{h}_{A}^{\ppe} \equiv A_{A}^{\ppe}(f) \exp[i \Psi_{A}^{\ppe}(f)],
\label{ppe-waveform-par}
\ee
where $A = ({\textrm{I}},{\textrm{M}},{\textrm{RD}})$ for the inspiral, merger and ringdown phases
respectively.
Then, we shall put these waveforms together to construct a ppE waveform for the
entire coalescence. 
We conclude this section with a discussion of the data analysis cost incurred due to the addition 
of ppE parameters to the waveform.

\subsection{Inspiral}

Let us employ the insight gained by the analysis of Sec.~\ref{Inspiral-Sec} to propose inspiral ppE waveforms.
Consider then the ppE GW phase, a generic form of which can be extrapolated from Eq.~\eqref{insp-GW-phase-gen}.
One simple parameterization for the test system is the following:
\be
\Psi_{\I}^{\ppe,1}(f) = -\frac{\pi}{4} - \phi_{c} + 2 \pi f t_{0} + \beta u^{b}
+ u^{-5/3} \sum_{k=0}^{7} \psi_{k}^{\PN} u^{k/3},
\label{Psi-param}
\ee
where $\psi_{k}^{\PN}$ are PN phase coefficients, some of which are given in Eq.~\eqref{PN-phase-par}, 
while $\beta$ and $b$ are ppE parameters.
Such a parameterization allows one to easily map between BD GW modifications and massive graviton ones, 
through the parameter choices $(\beta,b) = (\beta_{\BD},-7/3)$ and $(\beta,b) = (\beta_{\MG},-1)$ respectively. 
 
One generalization would be to allow more than one alternative theory to be modeled simultaneously 
by the ppE waveform. With the test system in mind, such an idea suggests 
\ba
\Psi_{\I}^{\ppe,2}(f) &=& -\frac{\pi}{4} - \phi_{c} + 2 \pi f t_{c} + \gamma u^{-7/3} + \delta \; u^{-1} 
\nonumber \\
&+& \beta u^{b}
+ u^{-5/3} \sum_{k=0}^{7} \psi_{k}^{\PN} u^{k/3}.
\label{Psi-2}
\ea
The benefits of this parameterization is that it explicitly incorporates the GW phases predicted by
alternative theories, as well as the PN prediction, while allowing, through the $u^{b}$ term, for new alternative theory modifications that have not yet been developed. 
Note however that now we have introduced a degeneracy into the waveforms; {\emph{i.e.}}~when $b=-7/3$, waveforms along
the line $\gamma+\beta={\rm const.}$ are identical. In a sense then this is a non-optimal
expansion.

A better suggestion might then be to postulate
\be
\Psi_{\I}^{\ppe,3}(f) = -\frac{\pi}{4} - \phi_{c} + 2 \pi f t_{c} + u^{-7/3} \sum_{k=0}^{9} \phi_{k} \; u^{k/3}.
\label{Psi-3}
\ee
The controlling factor in the phase expansion has been modified to $-7/3$ instead of $-5/3$ to allow for dipole radiation. 
The phase parameters $\phi_{k}$ can take on any values; the ppE-GR templates will be those
where $\phi_{k}=\psi_{k}^{\PN}$.
Continuing in this vain then, the most general expansion we propose here for the
phase is the following:
\be
\Psi_{\I}^{\ppe,4}(f) = -\frac{\pi}{4} - \phi_{c} + 2 \pi f t_{c} + \sum_{k=0}^{N-1} \phi_{k} \; u^{b_{k}},
\label{Psi-4}
\ee
where now $\phi_{k}$ and $b_{k}$ are each a set of $N$ ppE parameters.
The ppE-GR templates are those with $\phi_{k}=\psi_{k}^{\PN}$, and
$b_{k}=(k-5)/3$. Again, this family can also
capture the inspiral phase of several well known alternative theories (see Appendix A)
with appropriate choices of the parameters.
The parameterizations in $\Psi_{\I}^{\ppe,3,4}$ are a generalization of the studies in~\cite{Arun:2006yw,Arun:2006hn,2009arXiv0907.2186C}, 
except that we here allow for non-GR exponents in the u-dependence of the phase that could model dynamical strong-field corrections to GR. 

Now let us concentrate on amplitude parameterizations. A generic form for the amplitude that can be extrapolated from 
Eq.~\eqref{insp-GW-amp-gen} is the following
\be
A_{\I}^{\ppe}(f) = 
-\left(\frac{5}{24}\right)^{1/2}  \pi^{-2/3} \frac{{\cal{M}}^{5/6}}{D_{L}}
Q f^{-7/6} \left[1 + \alpha(\lambda_{\GR}^{\mu}) u^{a}\right].
\label{A-param}
\ee
The quantities $\alpha(\lambda_{\GR}^{\mu})$ and $a$ are a ppE function and parameter respectively, 
where the former could depend on the 
GR parameter vector $\lambda^{\mu}_{\GR} = \left[{\cal{M}},\iota,F_{+,\times})\right]$. 
Such a non-GR deformation of the GW amplitude is not degenerate with high-order PN 
corrections, since these scale with higher-harmonics of the GW signal, instead of the fundamental mode. 
Moreover, such a parameterization reduces to well-known alternative theory predictions, such as CS modified gravity when 
$a = 1$ and $\alpha$ reduces to the second term in Eq.~\eqref{resp-CS}.
Of course, one could in principle introduce $N$ terms to the GW amplitude with different frequency exponents,
though for simplicity we will not do so here; furthermore, GW detectors are most sensitive to phase, so
there might be less to gain by introducing expansive amplitude modifications.

\subsection{Merger and Ringdown}

Here we employ the considerations discussed in Sec.~\ref{Merger-RD-Sec} to construct a ppE waveform 
that models the merger and ringdown 
phase in a non-GR, model-independent fashion. For the merger phase, 
we have (see Eq.~\eqref{merger-wave} with Eq.~\eqref{ppe-waveform-par})
\be
A_{\M}^{\ppe}(f) = \gamma u^{c},
\qquad 
\Psi_{\M}^{\ppe}(f) = \delta + \epsilon u.
\label{merger-amp-phase}
\ee
The ppE parameters during the merger are then the real numbers $(c,\epsilon)$, while the parameters $(\gamma,\delta)$ are
set by requiring continuity with the inspiral phase. 
When the ppE parameters are $(c,\epsilon) = (-2/3,2 \bar{t}_{0} {\cal{M}}^{-1})$ one recovers the GR prediction for the test system. 

A simple proposal for a ppE parameterization of the ringdown waveform is
\be
A_{\RD}^{\ppe,1}(f) = \zeta \frac{\tau}{1 + 4 \pi^{2} \tau^{2} \kappa \left(f - f_{RD}\right)^{d}},
\label{fitting}
\ee
with $\Psi_{\RD}^{\ppe} = 0$, and where $f_{\RD}$ and $\tau$ are the dominant quasi-normal (QN) frequency and decay time,
$\kappa$ is a real ppE function of the final mass of the remnant, 
$d$ is a real ppE parameter, the quantity
$\zeta$ is determined by continuity with the merger phase, 
and where we have chosen the transition to occur at the fundamental ringdown 
mode $f_{\MRD} = f_{\RD}$. This models
the ringdown with a single decaying mode, although even in GR there are an infinite number of such 
modes excited during ringdown (a possibility we analyze below). 

A more general ringdown waveform is as follows~\cite{Ferrari:2000ep}
\be
A_{\RD}^{\ppe,2}(f) = \sum_{\ell,m,n} \zeta \frac{\tau_{\ell m n}}{1 + 4 \pi^{2} \tau^{2}_{\ell m n} \kappa \left(f -\bar{f}_{\ell m n}\right)^{d}},
\label{RD-waveform}
\ee
where $\bar{f}_{\ell m n}$ and $\tau_{\ell m n}$ are the frequency and damping time of the $(n \ell m)$-th mode, while the ppE function $\kappa$
and parameter $d$ are assumed mode-independent. One could in principle postulate different ppE parameters for different 
modes, but this would introduce an immense number of such parameters, not all of which would necessarily be independent. 

Traditionally, suggested tests of alternative theories of gravity using ringdown GWs are based on the ``no-hair'' theorem
of GR.
This theorem states that given the mass and angular momentum of a perturbed BH, 
all frequencies and damping times of the waveform are determined. For example, 
in GR the $(220)$ mode is given by~\cite{Echeverria:1989hg,Berti:2009kk}
\ba
M \omega_{220} & \sim & 1.5251 - 1.1568 \left(1 - a \right)^{0.1292},
\nonumber \\
Q_{220} & \sim & 0.700 + 1.4187 \left(1 - a \right)^{0.4990},
\ea
where the total mass of the remnant is $M$, the spin Kerr parameter of the remnant is $a \equiv J/M^{2}$, 
and we have defined the quality factor $Q_{\ell m n} \equiv \omega_{\ell m n}/(2 \tau_{\ell m n})$.
The standard no-hair test assumes that two modes have been detected, the first of which can be used to determine $a$ and $M$, 
while the second one can be used to check for consistency through equations similar to those above 
for additional modes.
The violation of the two-hair theorem is common in alternative models for compact objects~\cite{Berti:2006qt}.
With the parameterization of Eq.~\eqref{RD-waveform}, no assumption has been made about the relation between
different $\ell m n$ modes, and thus, such a test is still possible. 

One fact that should be kept in mind is that the strength of decay of the 
QNR modes ({\emph{i.e.}}~the scaling of $\tau_{\ell m n}$ with harmonic number) 
depends strongly on how ``exotic'' the compact object is. In GR, the first quasinormal mode is dominant 
because $\tau_{\ell m n}$ increases rapidly with $n$,$ \ell$ and $m$. This is not the case, for example, 
with boson stars~\cite{Berti:2006qt}; in such objects there are classes of modes
with comparable damping times, yielding a signal that cannot be extracted with one (or a small set) of modes. 
We shall not concern ourselves with this issue here, beyond pointing out that some strategies have been 
developed to deal with it~\cite{Berti:2005ys}.  

\subsection{The Complete ppE Waveform}
\label{com-waveform}

In summary then, the complete ppE waveform we have proposed takes on the 
following piece-wise continuous form
\ba
\tilde{h}(f) =
 \begin{cases}
  \tilde{h}_{\I}^{\ppe, A}(f) & \text{for $f < f_{\IM}$}, \\
  \tilde{h}_{\M}^{\ppe}(f) & \text{for $f_{\IM} < f < f_{\MRD}$}, \\
  \tilde{h}_{\RD}^{\ppe,B}(f) & \text{for $f > f_{\MRD}$},
\end{cases}
\ea
where $A$ is any of $A = [1,\ldots,4]$ and $B$ is any of $B = [1,2]$. All pieces of the waveform are decomposed into an amplitude and
a phase with the structure of Eq.~\eqref{ppe-waveform-par}. 
During the inspiral, the amplitude is given by Eq.~\eqref{A-param}, while
the phase can be any of Eq.~\eqref{Psi-param},~\eqref{Psi-2},~\eqref{Psi-3},~\eqref{Psi-4},
depending on the level of bias one wishes to allow for.
During the merger, the amplitude and phase are given by Eq.~\eqref{merger-amp-phase}, while in the ringdown the phase is zero
and the amplitude is given by either Eq.~\eqref{fitting} or~\eqref{RD-waveform} depending on the level of generality desired.
The quantities $f_{\IM}$ and $f_{\MRD}$ are the inspiral-merger and the merger-ringdown frequency transitions respectively.
These quantities can also be considered ppE parameters, or alternatively $f_{\IM}$ and
$f_{\MRD}$ can be set to the light-ring frequency and dominant ringdown mode of the corresponding
remnant ppE-GR BH, for example.

Note that these waveforms are not analytic; they are continuous but non-differentiable
at the patching frequencies between the phases. Of course, if desired, ppE families with
higher levels of differentiability could be constructed, though our main goal here
was to introduce the ppE framework with a simple (yet potentially useful and relevant) model.

Furthermore, one could easily extend these waveforms with additional ppE parameters
to try to capture larger classes of deviations from GR. However,
there is an increasing price one must pay the more parameters are added to a fitting routine:
an increase in the number of filters needed
and a deterioration in the accuracy of parameter estimation.
The extra computational cost needed to deal with additional parameters could be prohibitively high.
For example, one can show that the introduction of one single extra parameter
to ringdown waveforms increases the number of required templates to cover the parameter space from $500$ to 
$10^{6}$ in a frequentist, matched-filtering scheme~\cite{Berti:2007zu}. 
In view of this, one must ensure that additional parameter included are truly well-motivated, 
as it is explained in the Appendices. 

At first glance, another potential problem with these ppE templates, were they to be used directly for GW
searches of equal mass, binary BH inspirals, is they do contain some modeling error, 
{\emph{i.e.}}~with all ppE parameters set to their GR values, the waveforms will not exactly
describe the merger event in GR. In principle one can construct ppE waveforms with as small
a modeling bias as desired, though this may not be practical, in particular if hybrid waveforms
using numerical solutions for the late inspiral and merger phase of the waveforms are needed.
The numerical waveform segments will only be known as sets of numbers---either direct samplings
of the simulation waveforms, or coefficients of some truncated series expansion fit to the latter---and
it is unclear how such waveforms could then be extended in any well-motivated, model independent manner as 
done here. 

However, as discussed in the Introduction, we do not propose to use the 
ppE-templates for direct detection, but rather in post-analysis {\em after} a detection has been made
with GR templates. Some modeling error is then not much of an issue, other than possibly
limiting the accuracy with which the given ppE family can measure deviations from GR.
Also, there are alternative techniques to ``brute-force'' matched filtering that would
alleviate some of the problems with increased numbers of parameters, though
we leave further study of this in the context of the ppE framework to future investigations~\cite{vecchio-new}.

\section{Conclusions}
\label{conclusions}

In this paper we have considered the notion of fundamental bias in GW astrophysics. 
The main source of this bias is the assumption that GR correctly describes GW physics throughout
the generation of the waves, and their subsequent propagation.
If GR is wrong in the dynamical strong-field regime, a region yet to be constrained by
observational tests, and templates based solely on GR are used,
such a bias could cause wrong inferences to be made about the population and nature of
corresponding sources in the universe.

We have introduced the {\em parameterized post-Einsteinian}, or ppE framework to 
help remedy this problem of fundamental bias for classes of sources that do 
not deviate too radically from GR to prevent {\em detection} with GR templates. 
The ppE framework introduces non-GR parameters in the waveform to model
non-GR deviations, in analogy to the ppN framework developed for Solar System tests. 
As an example, we focused on one source for which there is good observational
evidence that GR correctly describes (most of) the event: the quasi-circular inspiral,
merger and ringdown of equal-mass, non-spinning BH-like objects.
Each ppE parameter is ostensibly selected to characterize a violation of some fundamental GR 
property or principle, for example emission of dipole radiation; we were furthermore careful to
choose parameters so that the ppE template family would be sensitive to known effects of certain
theories of gravity, as reviewed in Appendix A.

The acknowledgement of bias and the formal introduction of the ppE framework opens up an entire new set of possible studies,
as discussed at the end of Sec.~\ref{sec_alg}. Depending on whether the signal is a GR or non-GR one and whether the template 
is a GR or ppE one, one can attempt to answer several important questions that are critical to data analysis. For example, with a non-GR signal
and a GR template, one can determine how much systematic error induced by fundamental bias might contaminate
parameter estimation and perhaps even signal extraction (depending on SNR). Alternatively, given a non-GR signal
and a ppE template one can attempt to determine how well GW observations can truly constrain generic GR deviations. 
Through a systematic refinement of the ppE framework and the development of different ppE template families
appropriate for different sources (spinning BH binary coalescences, extreme-mass and intermediate-mass ratio inspirals, etc.), 
we shall thus be able to determine the true potential of GW observations to constrain our standard model of the gravitational interaction. 

There are numerous additional avenues to be explored here. Several include 
studying the systematic errors fundamental bias could introduce on 
inferences about a population of detections made with pure GR templates ({\emph{i.e.}}~question (1) 
in Sec.~\ref{sec_two_qs}), designing ppE families for additional events
(larger classes of binary BH mergers, binary NS or NS/BH mergers, NS
oscillations, etc.), refining the proposed ppE detection pipeline discussed in the Introduction,,
and exploring how strongly particular alternative theories may be constrained using ppE templates.

In addition, it would be very useful to understand how the merger
phase in compact object coalescence differs in alternative theories. Such an understanding
can only be gained via numerical simulations of such merger events in alternative theories that possess well-posed initial
boundary value problems. This would give some indication of
whether a simple interpolation between inspiral and ringdown would suffice,
or if not, give guidance on how to model this phase in a ppE fashion.

The proposed ppE framework could also become more powerful if coupled to coincident electromagnetic detections,
for those events that also emit photons.
In such a case, some of the system parameters, for example the mass and the luminosity distance, could be
independently (electromagnetically) determined, thus reducing the full (system plus ppE) parameter space and allowing 
for a more accurate measurement of ppE parameters. Electromagnetic observations, however, might also be affected by some
of the GR modifications discussed here. A fruitful direction for future research would then be to map the ppE parameters
introduced here to modifications of electromagnetic observations.  

\section{Acknowledgements}

We thank Emanuele Berti, Vitor Cardoso, Neil Cornish, Jon Gair, Walter Del Pozo, Alberto Vecchio, and 
John Veitch for detailed comments and suggestions. Both authors acknowledge support from the NSF 
grant PHY-0745779, and FP acknowledges the support of the Alfred P. Sloan Foundation.

\appendix
\section{A Bestiary of Alternative Theories of Gravity} 
\label{bestiary}
In this appendix we shall provide a brief review of some well-known alternative
theories of gravity. We refer the reader to the extensive references presented
in this Appendix for further reading. 

\subsection{Scalar-Tensor Theories and Jordan-Brans-Dicke-Fierz Theory}
\label{ST-app}
  
Scalar-tensor theories modify the action by the inclusion of a scalar field $\psi$
that couples to the Einstein-Hilbert term with a field-dependent
coupling function $A(\varphi)$ and a potential $V(\varphi)$, namely
(see eg.~\cite{lrr-2006-3,Will:1993ns} for details):
\ba
S_{\ST} &=& \frac{1}{16 \pi G} \int d^{4}x \sqrt{-g} \left[ R - 2 \varphi^{,\mu} \varphi_{,\mu} - V(\varphi)\right]
\nonumber \\
&+& S_{\mat}[\psi_{\mat},A^{2}(\varphi) g_{\mu \nu}],
\label{ST-gen-action}
\ea
where commas stand for partial differentiation, 
$g$ is the determinant of the Einstein metric $g_{\mu \nu}$, $S_{\mat}$ is the action for
$\psi_{\mat}$, and we have neglected the cosmological constant term. The scalar-tensor
theory action as written in Eq.~\eqref{ST-gen-action} is sometimes said to be in the
Einstein frame~\cite{Faraoni:1998qx,Faraoni:1999hp}.

The action as written in Eq.~\eqref{ST-gen-action} might seem to define a non-metric theory, 
because the {\emph{external}} matter degrees of freedom $\psi_{\mat}$ couple both to the metric tensor 
and to the scalar field $\varphi$~\cite{lrr-2006-3}. As we explain below, however, scalar-tensor theories are indeed
metric as the scalar field acts only as a mediator that influences the metric, but it is the latter that determines the motion
of external matter fields. By performing the conformal transformation $\tilde{g}_{\mu \nu} = A^{2}(\varphi) g_{\mu \nu}$
we can map the above action into
\ba
S_{\ST} &=& \frac{1}{16 \pi G} \int d^{4}x \sqrt{-\tilde{g}} \left[ \phi R - \frac{\omega(\phi)}{\phi} \phi^{,\mu} \phi_{,\mu} - \phi^{2} V \right]
\nonumber \\
&+& S_{\mat}[\psi_{\mat},\tilde{g}_{\mu \nu}],
\label{ST-action}
\ea
which is clearly a metric theory. In Eq.~\eqref{ST-action}, the new field $\phi$ is defined via $\phi \equiv A^{-2}$,
the coupling field $\omega(\phi)$ is defined as $\omega \equiv (\alpha^{-2} - 3)/2$, where $\alpha \equiv A_{,\varphi}/A$.
Equation~\eqref{ST-action} is sometimes said to be in 
the Jordan frame~\cite{Faraoni:1998qx,Faraoni:1999hp}.

Variation of the action with respect to the metric and the scalar field leads to the modified field 
equations~\cite{Will:1993ns,Campanelli:1993sm}, which in the Jordan frame are
\ba
\square \phi &=& \frac{1}{3 + 2 \omega(\phi)} \left( 8 \pi T_{\mu \nu}^{\mat} - \frac{d \omega}{d\phi} \phi_{,\mu} \phi^{,\mu} \right),
\nonumber \\
G_{\mu \nu} &=& \frac{8 \pi}{\phi} T_{\mu \nu} + \frac{\omega}{\phi^{2}} \left(\phi_{\mu} \phi_{\nu} 
- \frac{1}{2} g_{\mu \nu} \phi^{,\sigma} \phi_{,\sigma} \right)
\nonumber \\
&+&
\frac{1}{\phi} \left( \phi_{,\mu \nu} - \tilde{g}_{\mu \nu} \square \phi \right),
\ea
where $\square = \tilde{g}^{\mu \nu} \partial_{\mu \nu}$ is the D'Alembertian operator
and $T_{\mu \nu}^{\mat}$ is the matter stress-energy tensor. Because of the form of the modified field
equations, scalar-tensor theories are sometimes thought of as modifying Newton's gravitational constant via 
$G \sim 1/\phi$. When the coupling $\omega(\phi) = \omega_{\BD}$ is constant, 
then the scalar-tensor theory reduces to BD theory~\cite{PhysRev.124.925}.

Scalar tensor theories satisfy all of the necessary criteria we discussed in the Introduction. First, scalar-tensor
theories can be written as purely metric theories, as shown above. Second, these theories reduce exactly
to GR in the limit $\omega_{\BD} \to \infty$, and thus, it passes all precision tests for a sufficiently large coupling.  
As for the dynamical strong-field behavior, one might expect the gradients of the scalar field to become large
close to merger, although scalar-tensor theories are not built to introduce high-order curvature corrections. 
Early studies of scalar-tensor theories in NSs suggest that the dynamical strong-field behavior can be greatly 
modified, a process sometimes called {\emph{spontaneous scalarization}}~\cite{Damour:1998jk,Sotani:2004rq}. 

The additional theoretical criteria discussed in the Introduction are also satisfied by scalar-tensor
theories. Such theories have been shown to be well-posed as an initial value 
problem~\cite{Salgado:2007ep,2007CQGra..24.5667L}. Moreover, BD theory has been shown 
to possess stable Schwarzschild-like BH solutions that pass all precision tests~\cite{Campanelli:1993sm} 
for sufficiently large coupling. Finally, scalar-tensor theories are well-motivated from the low-energy 
effective theory limit of string theory. More precisely, 
when one integrates out all the string quantum fluctuations, one finds that the higher-dimensional 
string theoretical action reduces to a local field theory similar to a scalar-tensor 
theory~\cite{Garay:1992ej,1985NuPhB.261....1F}. The mapping between scalar-tensor theories and 
string theory is $\phi = e^{-2 \psi}$, where $\psi$ is the dilaton field that couples to matter via 
$m(\psi) = \bar{m} \exp(\beta \psi)$, where $\bar{m}$ is a constant and $\beta$ mediates the dilatonic coupling. 
By requiring the Strong Equivalence Principle to hold, one can show that the action takes the form of Eq.~\eqref{ST-action} 
in the Jordan frame with $\omega = (1/\beta^2 - 12)/8$~\cite{Garay:1992ej,1985NuPhB.261....1F}.
 
Scalar-tensor theories can also be mapped to the recently popular $f(R)$ class of theories. In the latter, one modifies
the Einstein-Hilbert action by replacing the Ricci scalar by some functional of this quantity. Of course, such an action
defines an infinite class of theories as $f(R)$ is arbitrary. Such arbitrariness 
removes a certain amount of predictability from $f(R)$ theories. Moreover, one can show that $f(R)$ theories
are equivalent to BD theory with $\omega_{\BD} = 0$, via the mapping $\phi = df(R)/dR$ and 
$V(\phi) = R df(R)/dR - f(R)$~\cite{Chiba:2003ir,Sotiriou:2006hs}. Due to its reduced predictability and 
equivalence with BD theory, we shall not consider $f(R)$ theories further in this paper. 
  
The emission and propagation of GWs in scalar-tensor theories is different from that in GR. 
The main effect of BD theory is the emission of monopolar and dipolar energy, which 
modifies the GW flux formula via $\dot{E} = \dot{E}_{\BD} + \dot{E}_{\GR}$, where the dominant
BD correction is dipolar and of the form~\cite{Will:1994fb,Scharre:2001hn,Will:2004xi}
\be
\dot{E}_{\BD} = -\frac{2}{3} \frac{\eta^{2} M^{4}}{r_{12}^{4}} \frac{S^{2}}{\omega_{\BD}},
\ee
where we recall that $M=m_{1}+m_{2}$ is the total mass, $r_{12}$ is the binary orbital separation and $\eta = m_{1} m_{2}/M^{2}$ is the
symmetric mass ratio. The quantity $S$ is a function of the sensitivity of the compact objects, 
related to the self-gravitational binding energy per 
unit mass, which is determined by the equation of state. Such a modification arises because
BD theory also corrects the compact object's effective mass, causing the location of the center of gravitational 
binding energy to disagree with the center of inertial mass. 
The effective mass now depends on the internal structure of the bodies, thus violating
the Strong Equivalence Principle. Although for pure NS binaries or pure BH binaries this effect is either small or 
identically cancels,  it does not vanish for mixed (BH-NS) binaries.
 
The Fourier transform of the GW metric perturbation is greatly affected by the BD modification to the GW flux formula,
since this propagates into modifications to the evolution of the orbital radius and GW frequency. 
One can show that the transform of the response function (defined in Sec.~\ref{anatomy}) 
takes the form~\cite{Will:1994fb,Scharre:2001hn,Will:2004xi}
\be
\tilde{h}_{\BD} = \tilde{h}_{\GR} e^{i \phi_{\BD}},
\ee
where $\tilde{h}_{\GR}$ is the Fourier transform of the response function in GR.
The Brans-Dicke phase correction, $\phi_{\BD} = -  \beta_{\BD}  \; u^{-7/3}$,  
where we recall that $u=\pi {\cal{M}} f$ is the dimensionless frequency parameter, 
the chirp mass ${\cal{M}} \equiv \eta^{3/5} M$ and the Brans-Dicke phase parameter
\be
\beta_{\BD} \approx \frac{5}{3584}\frac{S^{2}}{\omega_{\BD}} \eta^{2/5},
\label{betaBD}
\ee
neglecting terms of ${\cal{O}}(\omega_{\BD}^{-2})$. Note that in the limit $\omega_{\BD} \to \infty$ one recovers GR. 
In the Solar System, Doppler tracking of the Cassini spacecraft has led to the $(2 \sigma)$ constraint 
$\omega_{\BD} >  4 \times 10^{4}$~\cite{Bertotti:2003rm}. 
Studies have suggested that a GW detection
could place constraints on $\omega_{\BD}$ that are one or two orders of magnitude larger than the Cassini 
one~\cite{Scharre:2001hn,Will:2004xi,Berti:2005qd,Berti:2004bd,Yagi:2009zm}.

GWs in scalar-tensor or BD theory are not restricted to having only two polarizations, namely, these theories
also allow for a breathing mode. 
GW detectors, however, are not directly sensitive to breathing
modes due to the geometry of the interferometer.
In spite of this, GW detectors will be indirectly sensitive to these modes, since they will carry
energy away from the binary and thus naturally lead to a modification of the energy balance law.
The correction to the GW amplitude, and thus the GW response function, however, 
is subdominant since it leads to ${\cal{O}}(1/\omega_{\BD})$ corrections to the GR prediction~\cite{Corda:2009gt}. 

Apart from the inspiral phase, the merger and ringdown phases are also modified in scalar-tensor theories. 
The ringdown phase has been analyzed in~\cite{Sotani:2005qx}, where it was found that the QNR frequencies
in NS mergers are shifted in such theories. The plunge and merger regime have not yet been studied, although one expects
the largest GW modifications to arise in this region. In fact,~\cite{Sotani:2004rq} has shown that dynamical strong-field 
gravity effects in scalar-tensor theories can greatly affect NS parameters and their oscillations. 
     
\subsection{Massive Graviton Theories}

Massive graviton theories is the name given to models in which the gravitational force is propagated by a massive
gauge boson, {\emph{i.e.~}}a graviton with mass $m_{g} \neq 0$ or Compton wavelength $\lambda_{g} \equiv h/(m_{g} c) < \infty$. 
In GR, GWs can be thought of as massless gravitons propagating at the speed of light.
If the gravitons possess a finite Compton wavelength however, GWs will travel at velocities less than the speed of light. 
When we refer to massive graviton theories in the main text of this paper, we really mean corrections to the propagation of 
GWs as induced by an effective graviton mass. 

On a classical effective level, several theoretical frameworks exist where GWs propagate at speeds less than that of 
light~\cite{Rosen:1974ua,Visser:1997hd,Bekenstein:2004ne}. 
One example is Rosen's bimetric theory~\cite{Rosen:1974ua}, where gravitons follow 
null geodesics of some fiducial metric $\eta_{\mu \nu}$, 
while light follows null geodesics of some other metric $g_{\mu \nu}$~\cite{lrr-2006-3,Will:1993ns}.
A similar, more recent example is Visser's massive graviton theory~\cite{Visser:1997hd}, where the graviton
is given a mass at the cost of introducing a non-dynamical background metric, {\emph{i.e.~}}a prior geometry. 
Another recent example is TeVeS theory~\cite{Bekenstein:2004ne}, 
where the introduction of a scalar and a vector field lead to subluminal GW propagation.

Apart from the above effective, phenomenological models, quantum gravitational theories exist that 
also suggest gravitons could have a mass, such as for example 
loop quantum cosmology (LQC)~\cite{Ashtekar:2003hd,Bojowald:2006da}.
In LQC, the dispersion relation of propagating tensor modes acquires holonomy corrections during loop quantization, 
which lead to a mass~\cite{Bojowald:2007cd} $m_{g} = \Delta^{-1/2} \gamma^{-1} (\rho/\rho_{c})$,
where $\gamma$ is the Barbero-Immirzi parameter, $\Delta$ is related to the area operator, 
and $\rho$ and $\rho_{c}$ are the total and critical energy density respectively. 
In string theory inspired {\emph{effective theories}}, the four-dimensional graviton might also acquire a mass
if there are compact, extra dimensions~\cite{Dvali:2000hr}.
Unfortunately, a complete action for such a string theoretical-inspired, massive graviton theory 
that remains covariant and free of the van Dam-Veltman-Zakharov (vDVZ) discontinuity 
(in the limit as $m_{g} \to 0$, the effective theory does not reduce to GR) remains 
elusive.

Massive graviton theories do not generally satisfy any of the criteria described in the Introduction. 
Some of these models, including the string-theory inspired ones~\cite{Dvali:2000hr,ArkaniHamed:2002sp},
are {\emph{not}} weak-field consistent due to the vDVZ discontinuity.
Moreover, most of these models unavoidingly need to postulate the existence of 
prior geometry, making them Lorentz violating and forcing them to violate the Strong Equivalence Principle.
Finally, the main effect of massive graviton  theories is to modify GW 
{\emph{propagation}}, which is by definition a weak-field effect. Needless to say, due to these problems,
only limited effort has gone into the study of stable solutions~\cite{Bebronne:2009mz} or Cauchy well-posedness, 
although the latter is unlikely to be generically satisfied.

In spite of such fundamental problems, the GW community has been interested in GW propagation tests of 
massive graviton theories because they hold the potential to constrain a formally quantum gravitational effect.
Such tests postulate the phenomenological, special-relativistic parameterization~\cite{Will:1997bb}:
\be
v_{g} = \left(1 - \frac{m_{g}^{2}}{E_{g}^{2}} \right)^{1/2} \sim 1 - \frac{m_{g}^{2}}{2E_{g}^{2}},
\label{vg}
\ee
where $E_{g}$ is the energy of the graviton, $v_{g}$ its speed and the second equality arises in the limit of $m_{g}/E \ll 1$. 
Such a modified dispersion relation leads to a modified relation between graviton time of emission and 
arrival~\cite{Will:1997bb}:
\be
\Delta t_{a} = \left(1 + z\right) \left[ \Delta t_{e} + \frac{D}{2 \lambda_{g}^{2}} \left(\frac{1}{f_{e}^{2}} + \frac{1}{f_{e}^{'2}} \right) \right],
\ee
where $f_{e}$ and $f_{e}'$ are the frequencies of the two gravitons at emission and 
$D/D_{L} \sim 1 - z + {\cal{O}}(z^{2})$ in the limit of low redshift~\cite{Will:1997bb}. 
Such a change in the time of arrival leads to a modified
Fourier transform of the response function~\cite{Will:1997bb}
\be
\tilde{h}_{\MG} = \tilde{h}_{\GR} e^{i \Psi_{\MG}},
\ee
where the massive graviton correction is $\Psi_{\MG} = -\beta_{\MG} u^{-1}$ with
\be
\beta_{\MG} = \frac{\pi^{2} D {\cal{M}}}{\lambda_{g}^{2} (1 + z)}.
\label{betaMG}
\ee
As before, ${\cal{M}} \equiv \eta^{3/5} M$ is the chirp mass, where $M$ is the total mass and $\eta$ is the symmetric mass ratio.

There are numerous suggestions on how GW detections could be used to place bounds on $\lambda_{g}$~\cite{Will:1997bb,Finn:2001qi,sutton:2002:bgm,Will:2004xi,Berti:2005qd,Berti:2004bd,Arun:2009pq,Stavridis:2009mb,Larson:1999kg,Cutler:2002ef}. 
Such tests could lead to constraints as strong as present weak-field bounds, the best of which is 
$m_{g} < 3.6 \times 10^{-25} \; {\textrm{eV}}$~\cite{Finn:2001qi} or equivalently
$\lambda_{g} > 3.4 \times 10^{15} \; {\textrm{km}}$~\cite{Baskaran:2008za} from pulsar timing observations. 

Another feature of a massive graviton that is usually neglected is its effect on Newton's second law.
A general property of massive graviton theories is the introduction of a Yukawa-like suppression of the gravitational 
potential~\cite{Will:1997bb}
\be
V_{\MG}(r) = \frac{M}{r} e^{-r/\lambda_{g}}.
\ee
Such a modification is sometimes referred to as a {\emph{fifth force}}, which has led 
to the Solar System bounds on $m_{g}$ cited above~\cite{Talmadge:1988qz}. A slightly more general parameterization of 
the above Yukawa-like correction to the Newtonian potential is
\be
V_{\MG}(r) = - \frac{M}{r} \left( 1 + \gamma_{\MG} \; e^{-r/\lambda_{g}} \right),
\ee
where $\gamma_{\MG}$ is an undetermined parameter. Gravitational theories with compactified extra 
dimensions give such corrections to the effective 4-dimensional gravitational potential, where $\gamma_{\MG}$ depends directly
on the compactification~\cite{Kehagias:1999my}. Regardless of the Yukawa parameterization, GWs will also 
sense this modified potential as a change in the frequency evolution, although the correction is likely suppressed
by the ratio of the binary separation to the graviton Compton wavelength.

One last important effect in massive graviton theories
is the appearance of three additional longitudinal modes 
(see eg.~\cite{Dilkes:2001av}). Such an effect seems unavoidable if the massive
graviton correction to the action is of Pauli-Fierz type~\cite{Dilkes:2001av}. Longitudinal modes arise due to 
the non-vanishing of the NP scalars $\Psi_{2}$ and $\Psi_{3}$, and can be associated with 
the presence of spin-$1$ particles. 

Under the influence of longitudinal GWs, a circular arrangement of test particles will be deformed along the direction of propagation 
(see~\cite{Will:1993ns} for a graphical representation), which could lead to significant biases in
GW astronomy. 
Consider, for example, a situation where a GW is observed by a network of detectors, such that the inclination angle can be extracted with some accuracy. If GWs were longitudinal instead of transverse, the estimation of the inclination angle could have considerable systematic errors.
If there are observed electromagnetic counterparts, for example if binary NS mergers cause gamma-ray bursts (GRB),
such a bias might be noticed with coincident detections. However, if only the GRB is observed, and its sky
location is used to search for a corresponding GW event, the latter could be missed if most of the energy
was emitted in a longitudinal wave not accounted for in the post-analysis.

\subsection{Chern-Simons Modified Gravity}
\label{CS-app}

CS modified gravity is a particular, non-minimally coupled, scalar tensor theory, 
inspired by string theory and loop quantum gravity.
This extension proposes the inclusion of a parity-violating Pontryagin density to the Einstein-Hilbert
action (see~\cite{Alexander:2009tp} for a complete review):
\ba
S_{CS} &=& \kappa \int d^{4}x \sqrt{-g} R + \frac{\alpha}{4} \int d^{4}x \sqrt{-g} \vartheta \pont 
\nonumber \\
&-& \frac{\beta}{2} \int d^{4} x \sqrt{-g} \left[ \left(\partial \vartheta \right)^{2} + 2 V(\vartheta) \right],
\label{CS-action}
\ea
where $\pont =  {\,^\ast\!}R^{\alpha}{}_{\beta}{}^{\gamma \delta} \,R^{\beta}{}_{\alpha \gamma \delta}$, the dual to the Riemann tensor is ${\,^\ast\!}R^{\alpha}{}_{\beta}{}^{\gamma \delta} = \epsilon^{\alpha}{}_{\beta \sigma \chi} R^{\sigma \chi \gamma \delta}/2$ and $\kappa^{-1} = 16 \pi G$. The quantities $\alpha$ and $\beta$ are coupling {\emph{constants}}, while the quantity $\vartheta$ is a coupling field, and together they determine the strength of the CS modification. Note that the Pontryagin density is a topological term in four-dimensions, and thus, it can usually be integrated by parts and removed from the action. In CS modified gravity, however, this density is coupled to a {\emph{non-constant}} field, which forces non-trivial correction upon integration by parts. A non-dynamical version of this theory exists, defined by Eq.~\eqref{CS-action} with $\beta = 0$, but this model has been shown to be severally overconstrained~\cite{Grumiller:2007rv,Yunes:2007ss,Yunes:2009hc}.

The CS modification is motivated by the requirement that any quantum gravitational theory be anomaly-free. In perturbative string theory of type I, IIB or Heterotic, the absence of such a term leads to Green-Schwarz anomalies upon compactification. The requirement of anomaly-cancellation thus necessitates the inclusion of such a CS term~\cite{Polchinski:1998rr}. Even in the non-linear regime, Ramond-Ramond scalars induce such a CS term for {\emph{all}} types of string theory due to duality symmetries~\cite{Polchinski:1998rr,Alexander:2004xd}   

A CS extension of GR can also be motivated by loop quantum gravity. 
Here, such terms arise upon the promotion of the Barbero-Immirzi parameter
to a field when coupled to fermions~\cite{Taveras:2008yf,Mercuri:2009zt}. Recently, an embedding of such a loop quantum 
gravity effect has been worked out in the context of supersymmetric string theory~\cite{Gates:2009pt}.    

Upon variation of the action (\ref{CS-action}) with respect to all degrees of freedom we find the equations of motion of the theory:
\ba
G_{\mu \nu} + \frac{\alpha}{\kappa} C_{\mu \nu} &=& \frac{1}{2 \kappa} \left(T_{\mu \nu}^{mat} + T_{\mu \nu}^{\vartheta} \right),
\nonumber \\
\beta \square \vartheta &=& \beta \frac{dV}{d\vartheta} - \frac{\alpha}{4} \pont,
\ea
where $T_{\mu \nu}^{mat}$ is the stress-energy for non-gravitational matter degrees of freedom, while $T_{\mu \nu}^{\vartheta}$ is the stress-energy tensor 
of the $\vartheta$ field, namely
\be
T_{\mu \nu}^{\vartheta} = \beta \left[ \partial_{\mu} \vartheta \partial_{\nu} \vartheta - \frac{1}{2} g_{\mu \nu} \left( \partial \vartheta \right)^{2} - g_{\mu \nu} V \right].
\ee
The CS correction is then mostly encoded in the C-tensor, which is defined as
\be
C^{\mu \nu} = \left(\partial_{\gamma} \vartheta \right) \epsilon^{\gamma \delta \epsilon (\mu} \nabla_{\epsilon} R^{\nu)}{}_{\delta} + \left(\nabla_{\gamma \delta} \vartheta\right) {\,^\ast\!}R^{\delta (\mu \nu) \gamma},
\ee
where $\epsilon^{\mu \nu \delta \gamma}$ is the Levi-Civita tensor and $\nabla_{\mu}$ is the covariant derivative associated with $g_{\mu \nu}$.

The main criteria discussed in the Introduction are satisfied by CS modified gravity. By construction, this GR extension
is a metric theory that reduces to GR in the weak-field limit for sufficiently small gradients of $\vartheta$. 
Also, as opposed to scalar-tensor theories, CS modified gravity can potentially have large, purely
strong-field deviations due to the presence of the Riemann squared
term in the action. 

Some of the additional theoretical criteria discussed in the Introduction are also satisfied by CS modified gravity. This GR
extension is well-motivated by the leading candidates for a quantum theory of gravity.
CS modified gravity admits the Schwarzschild solution, and a modified Kerr solution~\cite{Yunes:2009hc,Konno:2009kg,Sopuerta:2009iy},
although the stability of these solutions with a dynamical scalar field (with non-vanishing background value) have not yet been investigated. 
The well-posedness of the theory as a Cauchy problem is difficult to
determine and it has not yet been worked out, although there are arguments that suggests the theory is well-posed
at the linear level (for details on this see~\cite{Alexander:2009tp}).  

One of the main effects of the CS modification is to induce parity-violation in gravitational interactions. In the non-dynamical
framework (when the CS scalar is fixed {\emph{a priori}} and devoid of dynamics), parity violation can be searched for in the Solar System~\cite{Alexander:2007vt,Alexander:2007zg,Smith:2007jm} and with double binary pulsar observations~\cite{Yunes:2008ua} through frame-dragging modifications. The best current bound on non-dynamical CS modified gravity comes from the latter, leading to the constraint $\dot{\vartheta}^{-1} > 33 \; {\textrm{meV}}$. In the dynamical framework (when the CS scalar is consistently determined by its coupled evolution equation), the CS modification is essentially unconstrained, as the known corrections only appear in the strong-field
regime~\cite{Yunes:2009hc,Konno:2009kg}. Moreover, the effects of CS terms can be enhanced in this regime in the
presence of dense matter~\cite{Alexander:2008wi}.   

GW tests have been proposed to constrain CS modified gravity during binary inspirals~\cite{Alexander:2007:gwp,LigoBi,Yunes:2009hc,Sopuerta:2009iy}. Such tests can be divided into those that study GW propagation effects~\cite{Alexander:2007:gwp,LigoBi} and those that concentrate on GW generation effects~\cite{Yunes:2009hc,Sopuerta:2009iy}. 
GW propagation is modified because left- and right-polarized modes obey different evolution equations, 
leading to the solution~\cite{Alexander:2007:gwp} $h_{R,L} = h^{GR}_{R,L} \; e^{i \Psi_{CS}}$, 
where the CS phase correction is given by
\be
\Psi_{CS} = i \lambda_{R,L} \pi f  H_{0}  \int_{0}^{z} dz \left(1 + z\right)^{5/2} \left[ \frac{7}{2} \frac{d \vartheta}{dz} + \left(1 + z \right) \frac{d^{2} \vartheta}{dz^{2}} \right],
\ee
and where $\lambda_{R,L} = \pm 1$ depending on whether the GW is right or left-polarized. Note that the correction is purely imaginary and $k$-dependent, leading to a frequency-dependent amplitude correction. 
The Fourier transform of the response function of a GW detector is then modified as follows~\cite{LigoBi}:
\be
\tilde{h} \sim \tilde{h}^{GR} \left[1 - 4 i \Psi_{CS} \frac{F_{+}^{2} + F_{\times}^{2} \zeta \left(1 + \zeta^{2}\right)}{F_{+}^{2} \left(1 + \zeta^{2}\right)^{2} + 4 F_{\times}^{2} \zeta^{2}}\right] 
\label{resp-CS}
\ee
where $F_{+,\times}$ are the beam-pattern functions of the detector, $\zeta = \cos{\iota}$ with $\iota$ the inclination angle. In Eq.~\eqref{resp-CS}, we have effectively neglected terms quadratic in $\vartheta$, and thus, we have ignored the possibility of a CS-induced resonance present in the detector's sensitivity band~\cite{LigoBi}. 

CS modified gravity also changes the generation of GWs, directly affecting their phase and amplitude. Although these corrections are quadratic in $\vartheta$ during the early inspiral, they can be large during the late inspiral and merger of compact objects. Strong non-linearities source large $\vartheta$-gradients, which in turn emit energy-momentum. Such energy emission modifies the GR flux formula, and thus, the orbital frequency evolution, leading to phase corrections. Recent studies have performed simulations of EMRIs in this theory~\cite{Yunes:2009hc,Sopuerta:2009iy}, which confirm that dynamical strong-field effects can lead to significant accumulative dephasing 
relative to GR.

\subsection{Einstein-Aether Theory}

Einstein-Aether theory is a Lorentz-violating model where the Einstein-Hilbert action is enlarged into~\cite{Jacobson:2000xp,Mattingly:2001yd,Eling:2004dk,Jacobson:2007fh,Jacobson:2008aj}:
\be
S_{EA} = \kappa \int d^{4}x \sqrt{-g} \left(R + K^{\alpha \beta}{}_{\mu \nu} \nabla_{\alpha} u^{\mu} \nabla_{\beta} u^{\nu} \right),
\ee
where again $\kappa^{-1} = 16 \pi G$,
\be
K^{\alpha \beta}{}_{\mu \nu} = c_{1} g^{\alpha \beta} g_{\mu \nu} + c_{2} \delta^{\alpha}_{\mu} \delta^{\beta}_{\nu} + c_{3} \delta^{\alpha}_{\nu} \delta^{\beta}_{\mu} + c_{4} u^{\alpha} u^{\beta} g_{\mu \nu},
\ee
and $c_{1,2,3,4}$ are coupling constants. Clearly, the action remains covariant, but it does not remain Lorentz invariant because the vector $u^{\mu}$, the ``aether,'' selects a preferred frame (eg.~that in which $u^{\mu}$ is at rest). This vector can be forced to have unit norm via an additional Lagrange-multiplier term in the action. 

The motivation to construct a theory that explicitly possess preferred-frame physics arises from quantum gravity. First, the existence of a unique vacuum in the quantum theory is believed to select a preferred frame~\cite{Jacobson:2008aj}. Second, exact Lorentz invariance is known to lead to quantum field theoretic divergences for states with arbitrarily high energy~\cite{Jacobson:2000xp}, which are usually eliminated through the introduction of a short-scale cut-off that breaks Lorentz invariance. Einstein-Aether theory is a model constructed to explicitly break Lorentz invariance, while preserving the general covariance of GR.   

Einstein-Aether theory satisfies all of the necessary criteria discussed in the Introduction. As is obvious from the form of the action, this model is a metric theory, which reduces to GR in the weak field for sufficiently small coupling constants $c_{1,2,3,4}$~\cite{Eling:2003rd,Foster:2005dk}. 
In the dynamical strong-field, however, it could disagree with GR, although the aether correction to the action is not constructed
from higher-order curvature terms. 
Some of the additional theoretical criteria discussed in the Introduction are also satisfied. Although this model is 
phenomenological, it is well-motivated by quantum field theory considerations. Stationary and stable BH solutions
have been found both analytically~\cite{Eling:2006ec} and numerically~\cite{Garfinkle:2007bk}. 
While the well-posedness of the Cauchy problem has not been studied separately, the theory has been shown to be 
well-posed numerically in certain dynamical scenarios~\cite{Garfinkle:2007bk}. 

Linearized Einstein-Aether theory possesses $5$ propagating degrees of freedom~\cite{Jacobson:2004ts}: one scalar (spin-0), three vectorial (spin-1) and two tensorial (spin-2) modes. The spin-0 mode leads to a breathing mode, while the spin-1 modes lead to longitudinal polarizations and the spin-2 ones to transverse-traceless or quadrupolar ones. The speed of propagation of these modes is found to be generically different from unity, where in particular for the spin-2 modes~\cite{Jacobson:2004ts}
\be
v_{g}^{2} = \frac{1}{1 - \left(c_{1}+c_{3} \right)}.
\ee
In spite of their apparent superluminal nature, the spin-2 modes possess positive-definite energies. 
Although Einstein-Aether theory predicts the existence of scalar and vectorial modes, these decouple in the weak-field limit and all radiation can be forced to be quadrupolar by some choice of the coupling constants~\cite{Jacobson:2008aj}. 

Compact objects are also modified in Einstein-Aether theory~\cite{Eling:2006df,Eling:2007xh,Eling:2006ec,Garfinkle:2007bk,Konoplya:2006rv,Konoplya:2006ar}. 
Non-rotating NSs have a slightly smaller radii relative to GR, which leads to larger redshift factors for electromagnetic
radiation produced at NS surfaces. The Roche lobes of interacting NS is also somewhat enlarged, leading to prompter tidal disruptions relative to the GR prediction. BHs possess slightly larger inner-most stable circular orbits (ISCO) relative to GR,
which shifts the GW signal to slightly larger frequencies during the late inspiral, plunge and merger~\cite{Eling:2006ec,Garfinkle:2007bk}.  

The evolution of binary systems is also modified in Einstein-Aether theory~\cite{Jacobson:2008aj,Konoplya:2006rv,Konoplya:2006ar}. Kepler's second law is modified in a manner that is dependent on the sensitivities of the bodies~\cite{Jacobson:2008aj}. Such a correction should be magnified during the late stage of coalescence, precisely when the system emits 
the most amount of energy in GW radiation. After the merger, the ringdown phase is also modified, since BHs possess
QNR frequencies larger than in GR~\cite{Konoplya:2006rv,Konoplya:2006ar}.

\subsection{MOND and TeVeS}

Modified Newtonian Dynamics (MOND) is a model proposed to explain galaxy rotation curves
without invoking cold dark matter~\cite{Bekenstein:1984tv}. This model was generalized to a 
relativistic theory through the postulate of a Tensor-Vector-Scalar (TeVeS) 
theory~\cite{Bekenstein:2004ne,Bekenstein:2005nv,Sanders:2006sz}. TeVeS and MOND have
managed to successfully explain a number of astrophysical phenomena beyond galaxy rotation curves, 
but we shall not discuss those here~(for a review, see~\cite{Sanders:2006sz}). 

The main modification introduced in MOND is the correction of Newton's second law via
\be
a^{i} \mu\left(\frac{|a^{i}|}{a_{0}}\right) = - \nabla^{i} \phi,
\ee
where $\phi$ is the Newtonian potential, $a_{0} \sim c H_{0} \sim 10^{-10} \; {\textrm{m s}}^{-2}$ 
is an intrinsic scale in the theory, 
and $\mu$ is a function that interpolates between $\mu(x) \sim x$ for $x \ll 1$ and $\mu \sim 1$ as $x \to \infty$. 
One popular choice for this function is $\mu = x (1 + x)^{-1}$,
which leads to the modified equation
\be
a^{i} = - \nabla^{i} \phi \left(1 + \frac{a_{0}}{|\nabla^{i} \phi|} \right),
\ee
where we have expanded in $|\nabla^{i} \phi| \gg a_{0}$ and we have assumed the zeroth order solution 
$a^{i} \sim - \nabla^{i} \phi$. 

How relevant is this modification to binary BH coalescences? The acceleration to compare 
$a_{0}$ to is $|\nabla^{i} \phi| \sim M/r_{12}^{2}$, where, as before, $M$ is the total mass and $r_{12}$ is the binary  
separation. The initial separation for a binary system with a period such that it is one-year from reaching $6 M$ is 
\be
r_{12} \approx 1.774 \times 10^{8} M_{\odot} \left(\frac{\cal{M}}{10^{6} M_{\odot}} \right)^{3/4} \left( \frac{0.25}{\eta} \right)^{1/5} \left(\frac{T}{1 \; {\textrm{yr}}} \right)^{1/4}.
\ee
Supermassive BH binaries (LISA sources) and NS binaries (ground-detector sources) can be sampled to initial separations of 
$\sim 10^{7} M_{\odot}$ and $\sim 300 M_{\odot}$ before coalescence respectively. For the most massive BH binaries $M \sim 10^{7} M_{\odot}$, with the largest initial separations $r_{12}=10^{8} M_{\odot}$, the Newtonian acceleration is very low, $10^{-3} \; {\textrm{m s}}^{-2}$, but still much larger than $a_{0}$. Based on this, MOND-like modifications to binary dynamics are negligibly small. 

A relativistic extension of MOND exists, namely TeVeS, which adds a dynamical vector and scalar field to the action,
making the theory stratified: the gradient of the scalar field and the vector field select ``strata'' or preferred spacetime sections.
This model is by definition {\emph{bimetric}}, where the physical metric differs from the auxiliary one, not only by a conformal factor
but also by terms that depend on the vector field. The interpolating function $\mu$ is here governed by the evolution of the scalar field, 
which itself is coupled to the evolution of the metric tensor. In view of this, one can think of TeVeS as a generic theory 
that interpolates between MOND in the slow-velocity limit and BD theory in the large-velocity limit~\cite{Sanders:2006sz}.  

Do these theories satisfy the necessary criteria discussed in the Introduction? As we mentioned above, TeVeS is 
a theory that depends on metric tensors only, and as such, it is a metric theory, albeit Lorentz violating. In this vein,
both MOND and TeVeS violate the Strong Equivalence Principle, due to the presence of preferred frames~\cite{Sagi:2009kd}. 
Weak-field consistency is actually violated since TeVeS reduces to MOND in this limit, instead of GR or 
regular Newtonian gravity, though of course if dark matter is due to an inaccurate description
of gravity on galactic scales, then consistency with Newtonian gravity here is not a desirable
feature. For TeVeS, in the dynamical strong-field one could expect large deviations from GR due to the presence of additional, dynamical scalar and vectorial fields. MOND is not well-motivated by fundamental 
theoretical considerations. TeVeS, being a covariant theory derivable from an action is arguably 
more well motivated, though it is unknown whether it admits a well-posed initial boundary-value
formulation~\cite{Bruneton:2007si}.

One consequence for GW detection in TEVES would be the modification of the balance law in binaries, since  additional scalar 
and vectorial degrees of freedom will also emit energy~\cite{Giannios:2005es}.
Another consequence is that TeVeS GWs propagate with speeds $v_{g} = c \; e^{-\phi} < c$, 
where $\phi$ is the TeVeS scalar field~\cite{Bekenstein:2004ne}. In regions of higher curvature (near singularities or close to binary mergers) $\phi$ 
is larger and thus GWs travel more slowly than in GR~\cite{Bekenstein:2004ne}.

\subsection{DGP Theory}

Several higher-dimensional theories have been proposed; one -- Dvali-Gabadadze-Porrati (DGP)~\cite{Dvali:2000hr,Dvali:2000rv}-- 
has received particular attention in the literature since it has the potential to explain the acceleration of the universe. 
Let us then summarize some of the principal features of DGP theory, following the review paper~\cite{Lue:2005ya}.

DGP is a metric theory, based on the string-inspired idea of braneworlds: higher-dimensional spaces of which our $4D$ universe
is only a surface, which for DGP is of co-dimension one. The extra dimension in DGP is assumed large and flat relative to astrophysical scales. 
As such, the theory introduces infrared modifications to the field equations, as opposed to ultraviolet ones, via leakage of gravitons into the 
extra dimension.  

By construction then, DGP gravity will not modify GR in the dynamical strong-field regions of relevance
to compact object GW sources.
Nonetheless, it is informative to study how such modifications could modify GWs emitted by cosmological binaries. 
Moreover, although the theory is somewhat well-motivated from a fundamental physics standpoint and although 
BH and weak-field, Solar System solutions exist, it remains unclear whether Cauchy well-posedness is satisfied due to the potential emergence of
ghosts or tachyon-like fields by the extra dimension~\cite{Charmousis:2006pn}. 

DGP has not really been studied in the context of GWs produced by binary systems. The only studies on GWs and DGP refer to
either cosmological tensor perturbations or shock waves~\cite{Kaloper:2005az,Kaloper:2005wa,Charmousis:2006pn}, 
neither of which are applicable for our purposes. Such a lack of interest in GW solutions is surprising, considering the DGP action 
is well-known, in vacuum reducing to
\be
S_{DGP} = - \frac{1}{16 \pi} \int  d^{5}x \sqrt{-g^{(5)}} \left(M^{3} R^{(5)} + M_{p}^{2} R^{(4)}\right),
\ee
where $M$ and $M_{p}$ are the five-dimensional and observed, four-dimensional Planck scale, 
$R^{(5)}$ is the five-dimensional Ricci scalar, $g^{(4,5)}$ are the determinants
of the induced-four and five-dimensional metric tensors and ${\cal{L}}_{m}$ is some additional matter lagrangian that couples to the four-dimensional
sector only. Note that we have reinstated the factors of $G$ here (through the explicit appearance of the Planck mass), 
as these get modified in DGP theory. With such an action, the DGP field equations become, in vacuum 
\be
M^{3} G_{AB}^{(5)} + M_{p}^{2} \delta\left[x^{5} - z(x^{\mu})\right] G_{AB}^{(4)} = 0,
\ee
where $A,B,\ldots = \{0,1,2,3,4\}$ stand for bulk indices and $\mu, \nu, \ldots = \{0,1,2,3\}$ for spacetime indices, with $x^{5}$ the extra dimension,
$z(x^{\mu})$ the location of the brane and $G_{AB}^{(4,5)}$ the four- and five-dimensional Einstein tensor.

One can now linearize the field equations about Minkowski spacetime to find that the spacetime components of the field equations become
(in Lorentz gauge):
\be
M_{p}^{2} \delta(x^{5})\square^{(4)} h_{\mu \nu}  =M_{p}^{2} \delta(x^{5}) h^{5}{}_{5,\mu \nu} - M^{3} \square^{(5)} h_{\mu \nu},
\ee
where $\square^{(4,5)}$ are the four- and five-dimensional D'Alembertian operators in flat space and we have chosen the brane to be located at
$x^{5} = 0$. If we now re-introduce matter fields, which couple to the four-dimensional sector of the field equations minimally, we can compute
the Fourier transform of the above equation, whose solution becomes
\be
\tilde{h}_{\mu \nu} = \frac{8 \pi}{M_{p}^{2} p^{2} + 2 M^{3} p} \left[\tilde{T}_{\mu \nu} - \frac{1}{3} \eta_{\mu \nu} \tilde{T}^{\alpha}{}{\alpha} \right].
\ee

The propagation of GWs is then modified at scales larger than $r_{0} \equiv M_{p}^{2}/(2 M^{3})$, which correspond to cosmological scales.
Alternatively, one can think of DGP modifications as introducing a spacetime-dependent Newton constant. For example, for scales $r < r_{0}$
the gravitational potential of an isolated point source of mass $m$ is
\ba
V(r) =
 \begin{cases}
  - G_{\textrm{brane}} m \; r^{-1} & \text{for $r \ll r_{0}$}, \\
  - G_{\textrm{bulk}} m \; r^{-2} & \text{for $r \gg r_{0}$},
\end{cases}
\ea
where $G_{brane} = M_{p}^{2}$ and $G_{bulk} = M^{-3}$. One sees that such a correction to the binding energy can be mapped to the
model-independent corrections of Sec.~\ref{mod-bin-ham} via $p = 1$. However, the cross-over scale here is gigantic, as since
$M \sim 100 {\textrm{MeV}}$ then $r_{0} \sim 10^{20} M_{\odot}$. Of course, if $M$ is larger, then the cross-over scale becomes smaller,
but then DGP would introduce ultraviolet corrections instead of infrared ones.

An interesting DGP correction to the GW response function would occur in the amplitude via the luminosity distance. Such a correction arises
because DGP introduces modifications to the Friedmann equations and its solutions. In particular, the Hubble parameter is modified in DGP
theory via
\be
H(z) \sim \frac{1}{2} H_{0} \left[ \frac{1}{r_{0} H_{0}} + \sqrt{\frac{1}{r_{0}^{2} H_{0}^{2}} + 4 \Omega^{0}_{M} ( 1 + z)^{3}} \right],
\ee
for a matter dominated and spatially flat cosmology, where $\Omega^{0}_{M}$ is the matter density content of the universe. Since the luminosity
distance depends on an integrated measure of the Hubble parameter, this quantity is modified in DGP theory. Such a correction could be 
recovered by the ppE parameter $\alpha$, which models amplitude deviations, for small redshift sources.

\section{Exotic Compact Objects} 
\label{exotic-BHs}

Many alternative theories of gravity have been discussed so far, but little has been said about the generic and model-independent possibility that a completely gravitationally collapsed object is not represented by the Schwarzschild or Kerr line elements. 
As discussed in the Introduction, though many observations of compact
objects consistent with BHs have been made, there is little to no strong evidence that they are indeed
BHs as described by GR (ie.~that they possess an event horizon).

Several alternatives exist that replace the Schwarzschild or Kerr singularities with other so-called 
{\emph{exotic}} alternatives. One such possibility arises from {\emph{q-balls}} 
(see eg.~\cite{Coleman:1985ki},~\cite{Kusenko:1998em}), namely a coherent scalar ``condensate'' that can 
be described classically as a non-topological soliton. These objects acquired some 
recent interest since they are unavoidable in viable supersymmetric extensions of the 
Standard Model~\cite{Kusenko:1997zq}. When the gravitational interaction is included, q-stars emerge; furthermore,
with gravity, a large class of scalar field matter sources admit similar stable solitons,
and these are often called boson stars 
(see eg.~~\cite{Friedberg:1986tq,Lynn:1988rb,Colpi:1986ye,Seidel:1991zh,Jetzer:1991jr,Verbin:2007hp}. Boson stars are fairly generic objects that could be present not just in GR, but also in other alternative theories of gravity, such as scalar-tensor ones~\cite{Balakrishna:1997ek}. Some interesting features of boson stars are, for example, angular momentum quantization in terms of the scalar charge~\cite{Kleihaus:2005me,Verbin:2007hp} and, that (even in 
pure GR) they can be almost as compact as BHs, though of course without an event horizon.

Horizonless, compact objects with large spins are problematic from a theoretical standpoint. In fact, boson stars, and all horizonless, compact objects that posses high spins have been shown to be unstable under small perturbations~\cite{Cardoso:2007az}. Such instability would lead to gravitational collapse to BHs, or possibly ``explosions''. Since many astrophysical BH candidates are believed to have high spins, such an instability restricts the interest of horizonless objects. Nonetheless, the existence of slowly-spinning or non-spinning, horizonless, compact objects cannot be ruled out by present observations.

Several works have looked at how merger dynamics might be different with boson stars, within the context of GR.
One study explored the merger of a small compact object with a supermassive boson star~\cite{Kesden:2004qx}, and
showed that stable orbits could exist inside the surface 
of the boson star. Such orbits exist because the effective potential for these geodesics inside the surface of the boson star
does not present the usual Schwarzschild-like singular behavior, but instead turns over and remains finite, allowing
for a new minimum. Such orbits lead to extreme precession that excite higher frequency harmonics of the waveform.

The merger and ringdown of boson star binaries is also drastically different from that of BH binaries. 
The merger of boson stars must be treated numerically; one such study explored a scenario where the merger
results in a single object
in a spinning bar configuration that either fragments or collapses into a 
Kerr BH~\cite{Palenzuela:2006wp,Palenzuela:2007dm}. 
The quasinormal ringing of boson stars is also different from that of BHs; furthermore the former do not satisfy the two-hair 
(or no-hair) theorem~\cite{Thorne:1997cw}. Instead, boson stars satisfy a three-hair theorem, where 
knowledge of the final mass, spin angular momentum {\emph{and}} quadrupole structure determines all successive 
QNR modes~\cite{Ryan:1996nk}. 
A different QNR spectrum in GW emission could be used to search for the existence of such boson stars~\cite{Ryan:1996nk,Thorne:1997cw,Dreyer:2003bv,Berti:2006qt}. 

Boson stars, however, are not the only exotic objects in the literature. Another interesting example is a gravitational vacuum star, or  {\emph{gravastars}}~\cite{Mazur:2001fv,Visser:2003ge}. In such objects, the event horizon is replaced by a phase transition that takes the outer Schwarzschild solution to an inner De-Sitter spacetime. Once more, the removal of an event horizon will lead to modified orbital dynamics, although a full analysis has not yet been carried out. Such stars will also lead to a different QNR spectrum~\cite{Chirenti:2007mk,Pani:2009ss}. 

Exotica is not limited to regular, horizonless alternatives;
for example, one could have BH solutions in GR that have exotic matter ``hair'', such as Skyrmion BHs~\cite{Shiiki:2005pb}. These ``dirty'' BHs might or might not cause modified orbital dynamics, but they will certainly lead to a different QNR spectrum~\cite{Medved:2003rga}.  
 
\bibliography{phyjabb,master}
\end{document}